\documentclass[]{jfm}

\usepackage{graphicx}
\usepackage[dvipsnames]{xcolor}
\usepackage{newtxtext}
\usepackage{newtxmath}
\usepackage{natbib}
\usepackage{hyperref}
\usepackage{physics}
\usepackage[retainorgcmds]{IEEEtrantools}
\usepackage{tikz}
\hypersetup{
    colorlinks = true,
    urlcolor   = blue,
    citecolor  = black,
}

\newcommand{\RomanNumeralCaps}[1]
\linenumbers

\newcommand{\beq}{\begin{equation}}
\newcommand{\eeq}{\end{equation}}
\newcommand{\bieee}{\begin{IEEEeqnarray}{rCl}}
\newcommand{\eieee}{\end{IEEEeqnarray}}

\newcommand{\nfft}{n_\mathrm{fft}}
\newcommand{\nblk}{n_\mathrm{blk}}

\title{Spectral dynamics of natural and forced supersonic twin-rectangular jet flow}

\author{Brandon Yeung\aff{1}
 \and Oliver T. Schmidt\aff{1}
  \corresp{\email{oschmidt@ucsd.edu}}}

\affiliation{\aff{1}Department of Mechanical and Aerospace Engineering, University of California San Diego, CA 92093, USA}

\begin{document}
\maketitle

\begin{abstract}
We study the stationary, intermittent, and nonlinear dynamics of natural and forced supersonic twin-rectangular turbulent jets using spectral modal decomposition. We decompose large-eddy simulation data into four reflectional symmetry components about the major and minor axes. In the natural jet, spectral proper orthogonal decomposition (SPOD) uncovers two resonant instabilities antisymmetric about the major axis. Known as screech tones, the more energetic of the two is symmetric about the minor axis and steady, while the other is intermittent. We test the hypothesis that flow symmetry can be leveraged for control design. Time-periodic forcing symmetric about the major and minor axes is implemented using a plasma actuation model, and succeeds in removing screech from a different symmetry component. We investigate the spectral peaks of the forced jet using an extension of bispectral mode decomposition (BMD), where the bispectrum is bounded by unity and which conditionally recovers the SPOD. We explain the appearance of harmonic peaks as three sets of triadic interactions between reflectional symmetries, forming an interconnected triad network. BMD modes of active triads distil coherent structures comprising multiple coupled instabilities, including Kelvin-Helmholtz, core, and guided-jet modes (G-JM). Downstream-propagating core modes can be symmetric or antisymmetric about the major axis, whereas upstream-propagating G-JM responsible for screech closure \citep{Edgington-MitchellEtAl2022JFM} are antisymmetric only. The dependence of G-JM on symmetry hence translates from the azimuthal symmetry of the round jet to the dihedral group symmetry of the twin-rectangular jet, and explains why the twin jet exhibits antisymmetric but not symmetric screech modes.

\end{abstract}


\section{Introduction}\label{sec:intro}
In turbulent jets, wavepackets are known to play an important role in noise generation due to their high spatiotemporal coherence and intermittency \citep{JordanColonius2013AnnuRev}. However, a complete and unified theory of the mechanisms that link wavepackets to observed noise remains elusive. The lack of such a theory is one factor that impedes the progress towards systematic control strategies for robust jet noise reduction.

Of particular interest in supersonic jet noise mitigation is the control of jet screech. \citet{Powell1953ASA} first identified this intense, spectrally-discrete acoustic phenomenon and proposed a resonant feedback loop as its origin \citep{Powell1953PSB}. This proposal has largely withstood the scrutiny of later research, even if details concerning the precise feedback mechanisms have evolved. It is generally accepted \citep{Edgington-Mitchell2019IJA} that the screech feedback loop is energised by the downstream-propagating Kelvin-Helmholtz (KH) instability of the turbulent mean flow \citep{Tam1971JFM}. The upstream-propagating waves that close the loop, however, are less well-understood, though a consensus is emerging \citep{ShenTam2002AIAAJ,GojonEtAl2018AIAAJ,Edgington-MitchellEtAl2018JFM,Edgington-MitchellEtAl2022JFM} that they are the subsonic instability waves, or guided-jet modes (G-JM), predicted by \citet{TamHu1989JFM}, which have also been shown to contribute to resonance in high subsonic jets \citep{TamAhuja1990JFM,TowneEtAl2017JFM,SchmidtEtAl2017JFM,JordanEtAl2018JFM,TowneEtAl2019AIAA}.

While most investigations of jet screech focus on the canonical round jet, there is increasing interest in more complex nozzle geometries, including the twin-rectangular jet that forms the subject of this study. The instability waves of round jets are well-characterised by the leading azimuthal Fourier modes \citep{JordanColonius2013AnnuRev}. By contrast, the rectangular nozzle shape and its associated mean flow support fundamentally distinct instabilities due to the absence of axisymmetry \citep{TamThies1993JFM,GutmarkGrinstein1999AnnuRev,RodriguezEtAl2021AIAA,NogueiraEtAl2023AIAA}. In addition, closely-spaced twin-rectangular jets create opportunities for the two jets to couple and interact \citep{RamanTaghavi1998JFM,KarnamEtAl2020AIAA,SamimyEtAl2023JFM,JeunEtAl2024JFM}. Like the rectangular, elliptical, and twin-round jets, the twin-rectangular jet possesses two axes of reflectional symmetry and thus belongs in the dihedral group $D_2$ \citep[see e.g.][]{SirovichPark1990POF}. Screech mitigation for the twin-rectangular jet must account for the unique instabilities that arise due to its $D_2$ symmetry. To this end, we carry out large-eddy simulations (LES) of a twin-rectangular jet flow using a nozzle geometry identical to the experimental set-up of \citet{SamimyEtAl2023JFM}, which was in turn derived from that of \citet{KarnamEtAl2020AIAA}. We show that twin-rectangular jet screech naturally adopts two classes of flapping instabilities. Based on this finding, we test---and subsequently confirm---the hypothesis that geometrical symmetries can be leveraged to control natural instabilities. 

Jet noise control has been investigated using a range of passive devices, including chevrons \citep{HeebEtAl2010AIAA,HendersonBridges2010AIAA,BridgesEtAl2011AIAA} and steady microjet injections \citep{GreskaEtAl2005AIAA}, as well as active devices. While many types of active devices, including plasma actuators, possess control authority in low-speed flows, the control of high-Reynolds number jets requires large-amplitude, high-bandwidth forcing \citep{CattafestaSheplak2011AnnuRev}. These requirements are the basis for the localised arc filament plasma actuators (LAFPA), developed by \citet{SamimyEtAl2004ExpFluids}. LAFPAs produce intense localised thermal perturbations through arc discharge. First experimentally demonstrated in round jets \citep{SamimyEtAl2004ExpFluids}, LAFPAs have recently been employed for noise control experiments in twin-rectangular jets \citep{SamimyEtAl2023JFM,SamimyEtAl2024JFM}. LAFPA-based control of single-rectangular jets has also recently been explored using LES, in which the actuators are modelled as boundary heating \citep{PrasadUnnikrishnan2023JFM,PrasadUnnikrishnan2024JFM}. We follow the approach pioneered by \citet{UtkinEtAl2007JPhysD} and later refined by \citet{KleinmanEtAl2009AIAA} and \citet{KimEtAl2009AIAA}, and model the actuators as volumetric energy sources. To maximise control authority, we fire the actuators in a pattern that respects the $D_2$ symmetry of the twin-rectangular jet.

Spectral proper orthogonal decomposition \citep[SPOD;][]{Lumley1970AP,TowneEtAl2018JFM,SchmidtColonius2020AIAAJ} and cyclostationary SPOD \citep[CS-SPOD;][]{HeidtColonius2024JFM} extract spatiotemporally coherent structures that optimally represent flow data in terms of energy. They have been used for physical discovery in natural \citep{SchmidtEtAl2018JFM} and highly forced jets \citep{HeidtColonius2024JFM}, respectively. Unlike SPOD, CS-SPOD is specialised for wide-sense cyclostationary flows, such as those that arise from periodic forcing, and discards the assumption of uncorrelatedness among frequencies. CS-SPOD targets triadic interactions involving the organised forcing and the stochastic turbulence, which only occur at very high forcing amplitudes. Instead, we are interested in the dynamics of the harmonic tones induced by the forcing. Using SPOD, we first examine the energy distribution of the natural and forced jets into both frequency and $D_2$ symmetry components. We show that screech is active in two of the four symmetry components of the natural jet. Forcing the jet in one of its inactive symmetry components leads to effective mitigation of screech in one of the two active components, but not the other.

Strong periodic actuation gives rise to harmonic peaks of the forcing frequency in the symmetry component of the forcing, but also to tonal peaks in a second symmetry component. These tonal peaks will be explained as inter-symmetry triad interactions.
In the present work, we are interested specifically in the occurrence or disappearance of tones in different symmetries of the forced jet. The occurrence of tones is facilitated by the nonlinearity of the Navier-Stokes equations, which establishes quadratic phase coupling between frequencies and symmetries. To distil the coherent structures that partake in nonlinear interactions, we use bispectral mode decomposition \citep[BMD;][]{Schmidt2020NDyn}. By detecting phase coupling between three frequency and symmetry components that are triadically compatible, BMD enables us to systematically catalogue dominant triads in the framework of bispectral statistics. BMD has been used to study frequency triads present in cylinder wakes \citep{Schmidt2020NDyn,FreemanEtAl2024JFM}, flat plate and aerofoil wakes \citep{Schmidt2020NDyn,DengEtAl2022POF,PatelYeh2023AIAA}, propeller and turbine wakes \citep{WangEtAl2023POF,KinjangiFoti2024TAML}, swirling flows \citep{SchmidtOberleithner2023JFM,FreemanEtAl2024JFM}, round impinging jets \citep{LiEtAl2024JFM,MaiaEtAl2024arXiv}, rectangular jets \citep{PrasadUnnikrishnan2024JFM}, and hypersonic boundary layers \citep{SousaEtAl2024ExpFluids}. Its utility for both frequency and symmetry triads has also been demonstrated on disk wakes \citep{NekkantiEtAl2023JFM} and round jets \citep{Schmidt2020NDyn,NekkantiEtAl2023AIAA}, which are continuously symmetric, as well as train wakes \citep{LiEtAl2024arXiv}, which are discretely symmetric. In the present study, we use BMD to extract structures associated with active frequency and $D_2$ symmetry triads.

The remainder of the paper is organised as follows. The numerical simulations and plasma actuation modelling are introduced in \S \ref{sec:method}. The decomposition of data into $D_2$ symmetry components is explained in \S \ref{sec:d2}. Section \ref{sec:energetics} presents spectral analysis of the natural and forced jets using SPOD. The nonlinear dynamics of the forced jet are explored using BMD in \S \ref{sec:nlin}. Results from SPOD and BMD are discussed in \S \ref{sec:discussion} and summarised in \S \ref{sec:conclusion}. Appendix \ref{sec:appMesh} reports on the computational grids used. Appendix \ref{sec:appBmd_recover_spod} details the recovery of SPOD from BMD. Appendix \ref{sec:appD2} outlines alternative approaches to the treatment of discrete spatial symmetries, such as $D_2$. Appendix \ref{sec:appNorm} compares the impact of different spatial norms on the modal statistics.

\section{Numerical set-up}\label{sec:method}
In this section we outline the set-up of the LES for the natural and forced jet simulations, and describe the modelling and implementation of plasma actuation in the forced jet.

\subsection{Large-eddy simulations}\label{sec:LES}
The simulations of the turbulent supersonic twin-rectangular jet are carried out using Cadence's unstructured, compressible LES solver `Charles' \citep{BresEtAl2017AIAAJ,BresLele2019RSTA}. The jet is nominally ideally-expanded and cold. The nozzle exit conditions are characterised by the jet Mach number, $M_\mathrm{j}=u_\mathrm{j}/c_\mathrm{j}=1.5$, acoustic Mach number, $M_\mathrm{a}=u_\mathrm{j}/c_\infty=1.25$, pressure, $p_\mathrm{j}/p_\infty=1$, temperature, $T_\mathrm{j}/T_\infty=0.69$, and Reynolds number, $\Rey_\mathrm{j}=\rho_\mathrm{j} u_\mathrm{j} D_\mathrm{e}/\mu_\mathrm{j}=1.07\times10^6$, where $u$ is the streamwise velocity, $c$ the speed of sound, $\rho$ the density, $D_\mathrm{e}/h=1.6$ the equivalent nozzle diameter, $h$ the nozzle height, $\mu$ the dynamic viscosity, and $(\cdot)_\mathrm{j}$ and $(\cdot)_\infty$ refer to jet exit and ambient conditions, respectively. The nozzle pressure (NPR) and temperature ratios (NTR) are $p_\mathrm{t}/p_\infty=3.671$ and $T_\mathrm{t}/T_\infty=1$, respectively. The biconical nozzle geometry, with an aspect ratio of two, is included in the computational domain and closely matches the experimental set-up of \citet{SamimyEtAl2023JFM}. The centre-to-centre spacing between the nozzles is $3.6h$. Each nozzle has a cavity cut into the internal wall just upstream of the nozzle exit. Housed within the cavity are eight pairs of electrodes per nozzle, allowing for up to eight plasma actuators per nozzle. Due to the sharp throat and cavity of the nozzles, shocks are present despite the jet being nominally ideally-expanded. Aided by the high Reynolds number, the cavity also facilitates a turbulent boundary layer state at the exit, captured using a wall model. The computational grid contains approximately 77 million cells. The simulation time step, $\dd{t}c_\infty/h$, is 0.002 for the natural jet, 0.001 for the forced jet (see \S \ref{sec:plasma_model}). These parameters are summarised in table~\ref{tab:LES_params}. For full details of the simulation, including its validation against the experiments of \citet{SamimyEtAl2023JFM}, we refer the reader to \citet{BresEtAl2021AIAA,BresEtAl2022AIAA}. The flow variables are non-dimensionalised by the ambient conditions, $\rho_\infty$, $T_\infty$, and $c_\infty=\sqrt{\gamma p_\infty/\rho_\infty}$, where $\gamma$ is the ratio of specific heats. Lengths are non-dimensionalised by $h$. Frequencies are non-dimensionalised by $u_\mathrm{j}/D_\mathrm{e}$, and reported as the Strouhal number, $fD_\mathrm{e}/u_\mathrm{j}$. For notational compactness, we denote frequencies by $f$, but emphasise that $f$ is dimensionless.

\begin{table}
\begin{center}
\def~{\hphantom{0}}
\begin{tabular}{lcccccccccc}
    Case & $M_\mathrm{j}$ & $M_\mathrm{a}$ & $\Rey_\mathrm{j}$ & $p_\mathrm{t}/p_\infty$ & $p_\mathrm{j}/p_\infty$ & $T_\mathrm{t}/T_\infty$ & $T_\mathrm{j}/T_\infty$ & $n_{\text{cells}}$ & $\dd{t}c_\infty/h$ \\[3pt]
    Natural & 1.5 & 1.25 & $1.07\times10^6$ & 3.671 & 1 & 1 & 0.69 & $76.6\times10^6$ & 0.002 \\
    Forced & 1.5 & 1.25 & $1.07\times10^6$ & 3.671 & 1 & 1 & 0.69 & $77.0\times10^6$ & 0.001 \\
\end{tabular}
\caption{LES parameters for the natural and forced jets.}
\label{tab:LES_params}
\end{center}
\end{table}

\subsection{Plasma actuation model}\label{sec:plasma_model}
To include the effect of plasma actuation in the LES, we adapt the model proposed by \citet{KimEtAl2009AIAA}. We insert a cylinder-shaped, volumetric source term into the right-hand side of the energy equation. The energy source is expressed as
\beq\label{eq:sourceTerm}
S(r,z,t) = k(t) l(r) m(z) P/(\pi r_0^2 L),
\eeq
where $P$ is the amplitude with the dimensions of power, $r_0$ is the radius of the cylinder, $L$ is its length, and $r$ and $z$ are local coordinates. For notational simplicity only, in equation~\eqref{eq:sourceTerm} we have aligned the local $z$-axis with the axis of the cylinder, such that $r=\sqrt{x^2+y^2}$. The exact geometry, including the locations and orientations of the actuators, can be found in \citet{SamimyEtAl2023JFM}. In actual implementation, the cylinder---like the plasma arc---is aligned with the nozzle lips. The envelope of the cylinder is given by the shape functions
\beq
l(r) = (1/2)[ \tanh(-\sigma(r/r_0-1/2)) +1]
\eeq
and
\beq
   m(z) = -(1/2)[\tanh{(-\sigma(z+L/2)/r_0)}+1] + (1/2)[\tanh{(-\sigma(z-L/2)/r_0)}+1],
\eeq
where $\sigma$ is a smoothness parameter. The energy source is modulated in time by the smoothed rectangular wave
\beq\label{eq:rectwave}
k(t) = (1/2)[ \tanh((t-n\tau-t_\mathrm{on})/t_\mathrm{r}) - \tanh((t-n\tau-t_\mathrm{off})/t_\mathrm{r}) ],
\eeq
where $t_\mathrm{on}$ and $t_\mathrm{off}$ are the time instants within each forcing period at which the signal is switched on and off, respectively, $t_\mathrm{r}$ is the rise time, $\tau$ is the forcing period, and $n=\lfloor t/\tau \rfloor$.

We assume $r_0$ to be half the depth of the cavity, and $L$ to be the distance between each pair of electrodes. The value for $\sigma$ is taken directly from \citet{KimEtAl2009AIAA}. The amplitude, $P$, and temporal parameters, $t_\mathrm{on}$, $t_\mathrm{off}$, and $t_\mathrm{r}$, are adapted from voltage and current measurements by \citet{SamimyEtAl2023JFM} of a typical actuation cycle. 
In general, these parameters can be set independently for each actuator; in this work, however, all actuators behave identically. The parameters are summarised in table~\ref{tab:plasma_params}. Based on our previous plasma modelling effort \citep{BresEtAl2021AIAA}, we apply additional grid refinement to the vicinity of the cavity, locally reducing the cell width by a factor of two relative to the natural jet (see appendix \ref{sec:appMesh}). The overall grid size is nearly unchanged. However, due to explicit time integration, the reduced minimum grid spacing necessitates a smaller time step (see table~\ref{tab:LES_params}).

Since it is challenging to obtain experimental measurements of the local flow environment around a plasma arc, rather than attempt to replicate a specific experiment, we seek to demonstrate the control authority of plasma actuation by exciting specific frequency and $D_2$ symmetry components in the twin-rectangular jet, as motivated in \S \ref{sec:intro}.

\subsubsection{Actuation strategy}
As we will see in \S \ref{sec:energetics}, the natural jet emits a dominant screech tone at a frequency of $f=0.29$. This screech is associated with flapping instabilities, i.e., antisymmetric oscillations about the major axis, $y=0$. To test the hypothesis that the natural antisymmetric instabilities may be disrupted, we force the twin jet symmetrically at the same frequency, $f_0=0.29$. Since the forcing specified by equation~\eqref{eq:rectwave} is a rectangular wave, the harmonics of the fundamental, $f_0,2f_0,\ldots$, are simultaneously excited. The experiments of \citet{SamimyEtAl2023JFM} employed six actuators per nozzle---three on the top lip, three on the bottom lip. We adopt the same strategy. Specifically, to achieve symmetric forcing, all 12 actuators will fire in phase at frequency $f_0$.
\begin{table}
\begin{center}
\def~{\hphantom{0}}
\begin{tabular}{lcccccccccc}
    $P/\rho_\infty c_\infty^3h^2$ & $r_0/h$ & $L/h$ & $\sigma$ & $t_\mathrm{on}c_\infty/h$ & $t_\mathrm{off}c_\infty/h$ & $t_\mathrm{r}c_\infty/h$ & $f_0$ \\[3pt]
    17.44 & 0.02 & 0.29 & 5 & 0 & 0.0015 & $2\times10^{-5}$ & 0.29 \\
\end{tabular}
\caption{Non-dimensionalised plasma actuator modelling parameters. Ambient temperature and pressure are assumed to be 293 K and 1 atm, respectively.}
\label{tab:plasma_params}
\end{center}
\end{table}

\begin{figure}
    \centering
    \includegraphics[width=\linewidth]{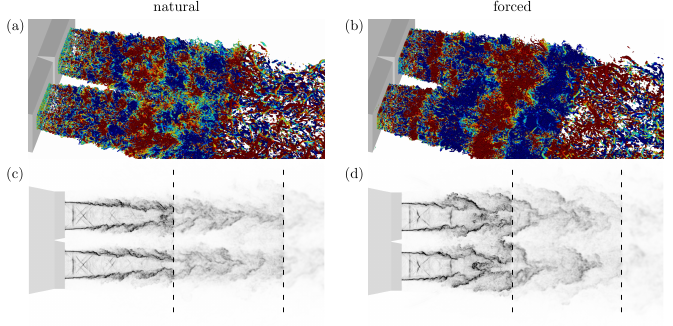}
    \caption{Instantaneous snapshots of the natural (a,c) and forced (b,d) jets: (a,b) $Q$-criterion isocontours of $Q=5$; (c,d) numerical schlieren, $|\grad\rho|$, on the major-axis plane, $y=0$. In (a,b), the contours are coloured by temperature fluctuations, $T'$. The colours saturate at $|T'|=\pm0.05$. In (c,d), the shading varies from white, $|\grad\rho|=0$, to black, $|\grad\rho|\ge10$. Dashed lines mark the locations of grid density transitions.}
    \label{fig:qcrit_schlieren}
\end{figure}
Instantaneous snapshots of the natural and forced jets are visualised in figure \ref{fig:qcrit_schlieren}. For both jets, the $Q$-criterion \citep{HuntEtAl1988CTR} isocontours display highly chaotic flow fields that contain a broad range of spatial scales. The forced jet shows alternating regions of high and low temperature due to the plasma actuation. In the numerical schlieren images, shock cells are observed within the potential cores. The shear layers of the forced jet reveal large-amplitude, symmetric perturbations as a result of the symmetric forcing pattern. Both the $Q$-criterion and schlieren visualisations show evidence of grid density transitions, as the LES filters fine-scale turbulence---a feature of LES subgrid models. These transitions occur at the boundaries between different grid refinement regions (see appendix \ref{sec:appMesh}), and are highlighted in figure \ref{fig:qcrit_schlieren}(c,d). However, as we will see later, the transitions have negligible impact on the large-scale structures we are interested in.

\section{Symmetries of the twin-rectangular jet flow}\label{sec:d2}
\begin{figure}
\centering
\begin{tikzpicture}
\draw (-.9,0) node [left] {$z$} -- (.9,0); \draw (0,-.25) -- (0,.25) node [above] {$y$};
\draw (-0.7,0) rectangle (-0.2,0.125); \draw (-0.7,0) rectangle (-0.2,-0.125);
\draw (0.7,0) rectangle (0.2,0.125); \draw (0.7,0) rectangle (0.2,-0.125);
\node [above] at (-.8,.25) {(a) SS};
\node [align=left] at (-1.625,0) {major\\axis};
\node [align=center,below] at (0,-.25) {minor\\axis};
\path [fill=lightgray] (2.5,0) rectangle (3.4,.25);
\path [fill=lightgray] (2.5,0) rectangle (3.4,-.25);
\draw (1.6,0) node [left] {$z$} -- (3.4,0); \draw (2.5,-.25) -- (2.5,.25) node [above] {$y$};
\draw (1.8,0) rectangle (2.3,0.125); \draw (1.8,0) rectangle (2.3,-0.125);
\draw (3.2,0) rectangle (2.7,0.125); \draw (3.2,0) rectangle (2.7,-0.125);
\node [above] at (1.7,.25) {(b) SA};
\path [fill=lightgray] (5,0) rectangle (4.1,-.25);
\path [fill=lightgray] (5,0) rectangle (5.9,-.25);
\draw (4.1,0) node [left] {$z$} -- (5.9,0); \draw (5,-.25) -- (5,.25) node [above] {$y$};
\draw (4.3,0) rectangle (4.8,0.125); \draw (4.3,0) rectangle (4.8,-0.125);
\draw (5.7,0) rectangle (5.2,0.125); \draw (5.7,0) rectangle (5.2,-0.125);
\node [above] at (4.2,.25) {(c) AS};
\path [fill=lightgray] (7.5,0) rectangle (6.6,-.25);
\path [fill=lightgray] (7.5,0) rectangle (8.4,.25);
\draw (6.6,0) node [left] {$z$} -- (8.4,0); \draw (7.5,-.25) -- (7.5,.25) node [above] {$y$};
\draw (6.8,0) rectangle (7.3,0.125); \draw (6.8,0) rectangle (7.3,-0.125);
\draw (8.2,0) rectangle (7.7,0.125); \draw (8.2,0) rectangle (7.7,-0.125);
\node [above] at (6.7,.25) {(d) AA};
\end{tikzpicture}
\caption{$D_2$ symmetry components. White and gray quadrants represent fluctuations of equal magnitude but opposite signs.}
\label{fig:d2_sketch}
\end{figure}
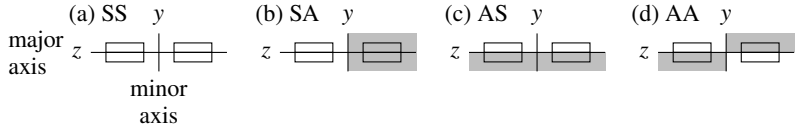

In the analysis of turbulent flows that enjoy statistical homogeneity in one or more spatial directions, it is customary to Fourier-transform the data along the homogeneous directions. Doing so reduces computational effort, accelerates the convergence of the statistics, and most importantly, enhances the interpretability of the results. For axisymmetric jets, this procedure amounts to an azimuthal Fourier transform. Twin jets, on the other hand, are inhomogeneous in all directions. Instead they possess $D_2$ symmetry, i.e., their geometries are invariant under reflection about the major and minor axes, $y=0$ and $z=0$, respectively. In such flows, $D_2$ symmetry gives rise to four symmetry components \citep{SirovichPark1990POF}: SS, SA, AS, and AA, where the first and second letters denote symmetry (S) or antisymmetry (A) about the major and minor axes, respectively \citep{RodriguezEtAl2018CRME}. These symmetry components are illustrated in figure \ref{fig:d2_sketch}.
We express the $D_2$ decomposition of pressure as
\beq\label{eq:d2Decomp}
p = p_\mathrm{SS} + p_\mathrm{SA} + p_\mathrm{AS} + p_\mathrm{AA}, 
\eeq
where the symmetry components are given by
\bieee\label{eq:d2PressureComponents}
p_\mathrm{SS} &=& \frac{1}{4}\qty[p(x,y,z,t) + p(x,-y,z,t) + p(x,y,-z,t) + p(x,-y,-z,t)] \IEEEeqnarraynumspace\IEEEyesnumber\IEEEyessubnumber\\
p_\mathrm{SA} &=& \frac{1}{4}\qty[p(x,y,z,t) + p(x,-y,z,t) - p(x,y,-z,t) - p(x,-y,-z,t)] \IEEEeqnarraynumspace\IEEEyessubnumber\\
p_\mathrm{AS} &=& \frac{1}{4}\qty[p(x,y,z,t) - p(x,-y,z,t) + p(x,y,-z,t) - p(x,-y,-z,t)] \IEEEeqnarraynumspace\IEEEyessubnumber\\
p_\mathrm{AA} &=& \frac{1}{4}\qty[p(x,y,z,t) - p(x,-y,z,t) - p(x,y,-z,t) + p(x,-y,-z,t)]. \IEEEeqnarraynumspace\IEEEyessubnumber
\eieee
As implied by equation~\eqref{eq:d2Decomp}, $D_2$ decomposition entails no loss of generality: the original flow field, $p$, can be exactly reconstructed by summing the four symmetry components in \eqref{eq:d2PressureComponents}. It is also possible to exploit statistical spatial symmetries by inflating the data record with geometrically-transformed copies of the snapshots \citep{Sirovich1987QAM2}. In appendix \ref{sec:appD2}, we outline this alternative procedure, and justify our choice to enforce symmetry via $D_2$ decomposition.

\begin{figure}
    \centering
    \includegraphics{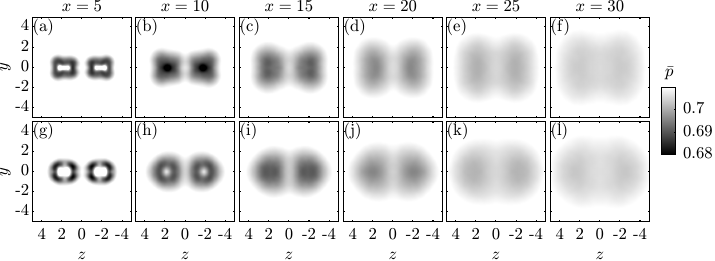}
    \caption{Long-time mean pressure of the natural (top row) and forced (bottom row) jets, visualised on axial planes: (a,g) $x=5$; (b,h) $x=10$; (c,i) $x=15$; (d,j) $x=20$; (e,k) $x=25$; (f,l) $x=30$. All panels share the same colour contours.}
    \label{fig:mean_p_x}
\end{figure}
To provide concrete motivation for $D_2$ decomposition, the long-time mean pressure of the natural jet is shown in figure \ref{fig:mean_p_x}(a--f) for different axial planes. On the $x=5$ plane, the mean profile bears overt resemblance to the twin-rectangular nozzle geometry. Much like the nozzle geometry, the mean is invariant with respect to reflections about the major and minor axes. Further downstream, the twin jets gradually begin to merge. The mixing of the twin jets remains incomplete even at $x=30$, where two distinct jet plumes are still visible. Over the entire domain of interest, $x\in\qty[0,30]$, the jet flow thus preserves its $D_2$ symmetry. This underscores the necessity of accounting for $D_2$ symmetry when analysing the statistics of the twin-rectangular jet.

The mean pressure of the forced jet, visualised in figure \ref{fig:mean_p_x}(g--l), undergoes an analogous streamwise evolution. The twin jets are separate for $x\lesssim5$, and partially mixed for $5\lesssim x\lesssim30$. The forced jet flow also remains $D_2$-symmetric for all $x$. However, the transverse mean profiles of the natural and forced jets reveal a stark difference, especially near the nozzle. Whereas the mean of the natural jet is shaped similar to two rounded rectangles, the mean of the forced jet resembles two ellipsoidal jets. Deformation of the mean flow by the forcing presages significant differences between the instabilities found in the natural and forced jets. We will begin examining some of these differences in \S \ref{sec:energetics} using SPOD.


\section{SPOD analysis}\label{sec:energetics}
SPOD has become the predominant data-driven method for the analysis of a broad range of turbulent flows, including turbulent jets \citep[see e.g.][]{GlauserEtAl1987TSF,SchmidtEtAl2018JFM,BresLele2019RSTA}. Its theory and implementation have been amply documented elsewhere \citep{Lumley1970AP,TowneEtAl2018JFM,SchmidtColonius2020AIAAJ}, so we will only briefly recapitulate the basics.

A time series made up of $n_t$ temporally- and spatially-discretised snapshots of a flow, $\vb*q_i$, $i=1,2,\ldots,n_t$, is segmented into $\nblk$ blocks of length $\nfft$ each, with an overlap of $n_\mathrm{ovlp}$ snapshots \citep{Welch1967IEEE}. A discrete Fourier transform (DFT) is applied to each block, yielding $\nblk$ Fourier realisations at each frequency, $\hat{\vb*q}^{(n)}_k$, $n=1,2,\ldots,\nblk$, $k=1,2,\ldots,\nfft$. For each frequency, the Fourier realisations are assembled in the data matrix,
\beq
\hat{\mathsfbi Q}_k = [ \hat{\vb*q}^{(1)}_k, \hat{\vb*q}^{(2)}_k, \cdots, \hat{\vb*q}^{(\nblk)}_k ].
\eeq
The SPOD modes, $\vb*\varPhi_k$, and modal energies, $\operatorname{diag}(\vb*\varLambda_k)$, are the solutions to the weighted eigenvalue problem
\beq\label{eq:spod_large_evp}
\mathsfbi S_k\mathsfbi W\vb*\varPhi_k = \vb*\varPhi_k\vb*\varLambda_k,
\eeq
where $\mathsfbi S_k = (1/\nblk)\hat{\mathsfbi Q}_k\hat{\mathsfbi Q}_k^*$ is the cross-spectral density (CSD) matrix and $(\cdot)^*$ denotes the conjugate transpose. The diagonal matrix $\mathsfbi W$ contains the numerical quadrature weights and ensures that the modes are optimal in the weighted 2-norm induced by the inner product $\expval{\vb*q_1,\vb*q_2}_{\vb*x}=\vb*q_2^*\mathsfbi W\vb*q_1$. For most flow data, equation~\eqref{eq:spod_large_evp} can be more efficiently solved via the method-of-snapshots \citep{Sirovich1987QAM1}, i.e.,
\beq\label{eq:spod_small_evp}
(1/\nblk)\hat{\mathsfbi Q}^*_k\mathsfbi W\hat{\mathsfbi Q}_k\vb*\varPsi_k = \vb*\varPsi_k\vb*\varLambda_k, \qq{with} \vb*\varPhi_k = (1/\sqrt{\nblk})\hat{\mathsfbi Q}_k\vb*\varPsi_k\vb*\varLambda_k^{-1/2}.
\eeq

\begin{table}
\begin{center}
\def~{\hphantom{0}}
\begin{tabular}{lccccccccccc}
    $n_x$ & $n_y$ & $n_z$ & $[x_\mathrm{min},x_\mathrm{max}]$ & $[y_\mathrm{min},y_\mathrm{max}]$ & $[z_\mathrm{min},z_\mathrm{max}]$ & $n_t$ & $\Delta tc_\infty/h$ & $\nfft$ & $n_\mathrm{ovlp}$ & $\nblk$ \\[3pt]
    130 & 60 & 76 & $[0,30]$ & $[-5,5]$ & $[-5,5]$ & 10 000 & 0.2 & 1024 & 512 & 18 \\
\end{tabular}
\caption{Database and spectral estimation parameters for SPOD (\S \ref{sec:energetics}) and BMD (\S \ref{sec:nlin}). The natural and forced jets share the same parameters.}
\label{tab:SPOD_params}
\end{center}
\end{table}
For each simulation (natural and forced), 10 000 snapshots of the LES are saved at a time interval of $\Delta tc_\infty/h=0.2$. Prior to statistical analysis, the unstructured snapshots are interpolated onto a Cartesian grid that spans $x\in[0,30]$, $y\in[-5,5]$, and $z\in[-5,5]$, and is discretised by $n_x=130$, $n_y=60$, and $n_z=76$ points in the $x$, $y$, and $z$ directions. The database interpolation and spectral estimation parameters are summarised in table~\ref{tab:SPOD_params}. In the remainder of this work, we will focus only on the pressure component of the data, but have confirmed in appendix \ref{sec:appNorm} that an analysis based on the primitive variables yields comparable results.

\subsection{SPOD modal energies and modes}
\begin{figure}
    \centering
    \includegraphics{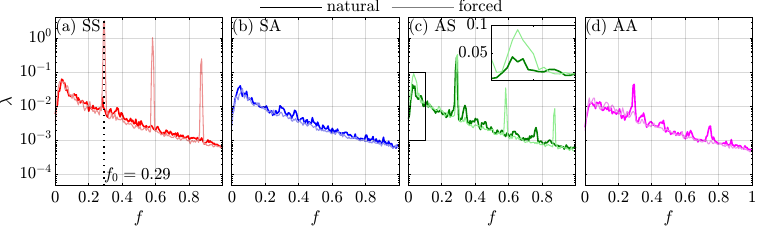}
    \caption{Leading SPOD eigenvalue spectra of the natural (solid lines) and forced (faded lines) jets: (a) SS; (b) SA; (c) AS; (d) AA symmetry components. The dotted line in (a) marks the SS forcing frequency, $f_0$. The inset in (c) zooms in on the frequency range $f\in\qty[0,0.1]$ and shows $\lambda$ on a linear scale.}
    \label{fig:spod_p_spec}
\end{figure}
The leading SPOD eigenvalue spectra of the natural and forced jets are reported and compared in figure \ref{fig:spod_p_spec}. In both cases, the modal energy distribution evidently depends on spatial symmetry. For the SS and SA symmetry components in figure \ref{fig:spod_p_spec}(a,b), the energy of the natural jet is predominantly broadband. Over the frequency range $f\gtrsim0.04$, the energy decays with frequency, as is expected in a turbulent flow. At very low frequencies, the energy rapidly falls off as $f\to0$ because the finite domain size limits the resolvability of structures with the longest wavelength (and thus lowest frequency). In the SS spectrum, a small peak can also be discerned at $f=0.28$. For the AS and AA symmetries in figure \ref{fig:spod_p_spec}(c,d), on the other hand, the spectra of the natural jet reveal a prominent tone at $f=0.29$ as a result of screech resonance. The screech tone is more energetic in the AS component than in AA. Screech tones are also present at higher frequencies, in particular at $f\approx0.34$ and 0.75, which are not harmonics of the dominant screech tone. They have substantially lower energy and stem from shock cells having non-uniform spacing \citep{Edgington-MitchellEtAl2022JFM}. Compared to the other three symmetries, the AA spectrum has significantly reduced energy at low frequencies, below $f\approx0.1$. Conversely, at low frequencies the SS symmetry is the most energetic. This is consistent with the strong in-phase coupling between the left and right jets observed at low frequencies by \citet{SamimyEtAl2023JFM}.

Whereas in the natural jet tones are found in the AS and AA symmetries, in the forced jet they have migrated to the SS and AS symmetries. These tones occur at the fundamental forcing frequency, $f_0=0.29$, as well as its harmonics. They make up part of the response of the jet to the exogenous forcing. Non-harmonic tones found in the AS spectrum of the natural jet are eliminated in the forced jet. The forced jet also responds by completely suppressing all tones in the AA component, producing a broadband spectrum. The SA spectrum remains broadband, and is nearly unchanged from the natural jet. Given the SS symmetry of the forcing, the generation of harmonics in the AS component and quelling of tones in AA are clear signs that nonlinear mechanisms are at work.

For the AS symmetry in figure \ref{fig:spod_p_spec}(c), a comparison between the natural and forced jets shows that the leading modes of both cases possess approximately equal energy at $f_0$. Below $f\approx0.1$, however, the two display distinct characteristics. As the inset in figure \ref{fig:spod_p_spec}(c) makes clear, for $0.01\lesssim f\lesssim0.07$, the forced jet has higher energy, suggesting it exhibits a greater prevalence of slowly-evolving structures. These changes are unsurprising considering the drastically modified mean flow shown in figure \ref{fig:mean_p_x}.
Below $f\approx0.01$, the separation between the natural and forced spectra initially shrinks, but diverges again as $f\to0$. In the $f=0$ frequency bin, the forced jet has tenfold the energy of the natural jet. By definition, the long-time mean cannot be spatially antisymmetric. As such, we attribute the energy of the AS component of the forced jet in the $f=0$ bin to coherent structures energised by the forcing that oscillate at frequencies below the finite bin width, $\Delta f=1/(n_\mathrm{fft}\Delta t)$. For the remaining three symmetry components, we observe no significant change to the low-frequency region of each of their spectra. We will revisit the modes at near-zero frequencies in \S\S \ref{sec:SSself}--\ref{sec:ASself_SSAS}.

\begin{figure}
    \centering
    \includegraphics{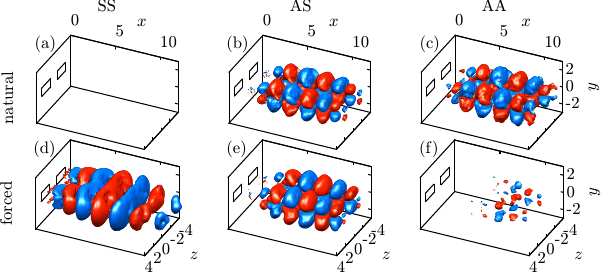}
    \caption{Leading SPOD modes, scaled by their SPOD amplitudes, of the natural (a--c) and forced (d--f) jets at $f=0.29$: (a,d) SS; (b,e) AS; (c,f) AA symmetry components. For each mode, isocontours of $\sqrt{\lambda_1}\Re{\phi_1(\vb*x)}=\pm d$ are shown in red and blue. The isovalue, $d$, is shared in each column: (a,d) $d=0.1$; (b,e) $d=0.05$; (c,f) $d=0.02$. Rectangles mark the exits of the twin nozzles.}
    \label{fig:spod_p_mode_3d}
\end{figure}
The leading SPOD modes of the natural and forced jets at frequency $f_0$ are shown in figure \ref{fig:spod_p_mode_3d}. By construction, SPOD modes are normalised such that they have unit energy, $\norm{\phi_1}^2_{\vb*x}=1$. To facilitate comparison of the relative presence of the modes as they occur in the data, we scale each mode by its SPOD amplitude, i.e., the square root of its corresponding mode energy, $\sqrt{\lambda_1}$. As the eigenvalue spectra in figure \ref{fig:spod_p_spec} reveal, only the SS, AS, and AA symmetry components of the natural or forced jet (or both) display tones. For this reason, we focus on the modes of these three symmetries in figure \ref{fig:spod_p_mode_3d}.

The 3D mode shapes clearly demonstrate the contrasting coupling behaviour of each symmetry component. For the SS symmetry, figure \ref{fig:spod_p_mode_3d}(a,d) shows the contours of the amplitude-scaled leading mode, $\sqrt{\lambda_1}\Re{\phi_1(\vb*x)}=\pm 0.1$, in red (positive) and blue (negative). At this isovalue, no structure is visible in the natural jet due to the low mode energy. Conversely, in the forced jet we observe coherent structures phase-locked to the strong SS excitation. The structures exit the twin nozzles independently, then rapidly begin to merge, so that by $x\approx5$ they have morphed into a single, large wavepacket that couples the left ($z>0$) and right ($z<0$) jets in phase with each other. For the AS symmetry, contours of $\sqrt{\lambda_1}\Re{\phi_1(\vb*x)}=\pm 0.05$ are shown in figure \ref{fig:spod_p_mode_3d}(b,e). Because the natural and forced jets possess almost equal energy at $f=0.29$ in this symmetry (as the spectra in figure \ref{fig:spod_p_spec}(c) indicate), the modes of both cases resemble each other. Like the SS modes, the AS modes represent in-phase coupling between the left and right jets. Unlike SS, however, the AS modes are antisymmetric about the major-axis plane, $y=0$. The AA modes in figure \ref{fig:spod_p_mode_3d}(c,f) display stark differences between the natural and forced cases. Whereas in the natural jet the contours of $\sqrt{\lambda_1}\Re{\phi_1(\vb*x)}=\pm 0.02$ reveal distinct structures, in the forced jet no discernible structure is present for the same isovalue. This is a direct consequence of the suppression of screech in the AA symmetry by the forcing. The envelope of the AA mode in the natural jet is qualitatively similar to the AS mode. That said, the former is distinguished by its antisymmetry about the minor-axis plane, $z=0$. The left and right jets are coupled perfectly out of phase, with a phase difference of $\pi$ between each other.

As we will see, SPOD and BMD modes of the twin-rectangular jet at the same frequency appear visually indistinguishable. Both methods successfully capture the most prevalent flow structures. We therefore defer a more detailed discussion of the physical mechanisms revealed by the modes until \S \ref{sec:nlin}.

\subsection{Intermittency}
\begin{figure}
    \centering
    \includegraphics{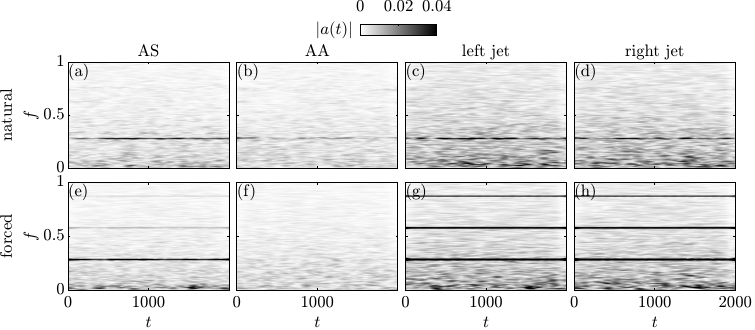}
    \caption{SPOD-based time-frequency analysis of the natural (a--d) and forced (e--h) jets: (a,b,e,f) AS and AA symmetry components; (c,d,g,h) the left and right jets primarily located in the $z>0$ and $z<0$ half domains, respectively. All panels share the same colour contours. 
    }
    \label{fig:tcoeffs}
\end{figure}
For the natural jet, the tonal peak at $f=0.29$ in the AS and AA symmetry components (see figure \ref{fig:spod_p_spec}) signifies the prevalence of flapping structures in the flow, but only in a statistical sense. In figure \ref{fig:tcoeffs} we investigate the temporal dynamics of these structures by conducting a time-frequency analysis of each symmetry. Specifically, we leverage the ability of the time-continuous SPOD expansion coefficients to provide insights into the global (as opposed to pointwise) evolution of modal structures \citep{SchmidtEtAl2017AIAA,TowneLiu2019APS}. For computational efficiency, we determine the expansion coefficients
\beq\label{eq:tcoeffs}
\mathsfbi A_k^{\qty{i}} = \vb*\varPhi_k^*\mathsfbi W\hat{\mathsfbi Q}_k^{\{i-\frac{\nfft}{2},\ldots,i+\frac{\nfft}{2}-1\}}
\eeq
for the time index $i$, with the Fourier transform $\hat{\mathsfbi Q}_k$ taken one snapshot at a time \citep{NekkantiSchmidt2021JFM}. The magnitudes of the complex-valued expansion coefficients are the time-dependent SPOD amplitudes. The coefficients corresponding to the leading mode are shown in figure \ref{fig:tcoeffs}(a,b) for the AS and AA symmetries, respectively, in the form of a time-frequency diagram. In figure \ref{fig:tcoeffs}(a), the AS symmetry displays a dark band at $f=0.29$, as expected from the dominant screech tone. The persistence of the band with only small fluctuations suggests the strength of the AS screech tone remains relatively stable over time. The frequency of the tone also remains fixed. On the other hand, the AA screech tone shown in figure \ref{fig:tcoeffs}(b) is intermittent. While its frequency is steady, its strength fluctuates significantly such that it vanishes, and later re-emerges, several times in the data record. Compared to the AA screech, the dominance and steadiness of the AS screech suggest the absence of competition between the two screech symmetries that would have otherwise led to symmetry-switching. This is in accordance with observations by \citet{WongEtAl2023JFM} that twin-round jets exhibit steady jet coupling at high NPR. The corresponding time-frequency diagrams for the forced jet are also shown in figure \ref{fig:tcoeffs}(e,f). As expected, the tonal peaks at frequency $f_0$ and its harmonics are clearly visible in the AS component, and do not manifest intermittency. No peak is visible in the AA component.

For completeness, we also perform independent time-frequency analyses of the left and right jets. For the left jet, SPOD modes are computed from the data in the $z>0$ half domain, without recourse to $D_2$ symmetry decomposition. The time-continuous expansion coefficients are obtained from equation~\eqref{eq:tcoeffs} as before. For the right jet, modes and coefficients are calculated in an analogous manner, just in the $z<0$ half plane. For the natural jet, the expansion coefficients of the left and right jets are reported in figure \ref{fig:tcoeffs}(c,d), respectively. The magnitudes of these coefficients are larger than those belonging to the AS and AA symmetries, because the left and right jets include all four symmetry components. 
Both the left and right jets appear intermittent, and lack a clear phase relationship between each other. The corresponding coefficients for the forced jet are displayed in figure \ref{fig:tcoeffs}(g,h). The harmonic peaks are visible and not intermittent, again as expected.

In light of the absence of symmetry-switching---as demonstrated by figure \ref{fig:tcoeffs}(a,b), and to simplify our analysis, in the remainder of this work we focus exclusively on results where $D_2$ symmetry is enforced.

\section{Nonlinear dynamics}\label{sec:nlin}





For the natural jet, the SPOD spectra in figure \ref{fig:spod_p_spec} do not manifest significant nonlinear activity. The spectra of the forced jet, on the other hand, uncover significant energy concentration and low-rankness at harmonics of the fundamental forcing frequency, $f_0,2f_0,\ldots$. This behaviour hints at the potential for nonlinear interactions among the harmonic frequencies in the forced jet. However, by construction, SPOD does not account for possible correlation between frequency components \citep{TowneEtAl2018JFM,HeidtColonius2024JFM}. In contrast, BMD \citep{Schmidt2020NDyn}, like the classical bispectrum on which it is based, is designed to discriminate between waves that are independently excited and thus uncorrelated, and those that maintain an approximately constant phase relationship over time and are statistically dependent \citep{KimPowers1979IEEE}. Application of BMD to the forced jet allows us to assess both the importance of nonlinear dynamics as well as how such dynamics may depend on frequency and spatial symmetry.

\subsection{Methodology}
\subsubsection{BMD with bicoherence normalisation}\label{sec:bicoherence}
For a one-dimensional, statistically stationary random signal, $q(t)$, and its Fourier transform, $\hat q(f)$, the classical bispectrum is defined as the triple correlation
\beq\label{eeq:classicBispectrum}
b(f_k,f_l) = \mathrm E\qty{\hat q^*(f_k)\hat q^*(f_l)\hat q(f_k+f_l)},
\eeq
where $\mathrm E\qty{\cdot}$ is the expectation operator. The frequency triplet, $(f_k,f_l,f_k+f_l)$, constitutes a triad. BMD generalises the classical bispectrum to flow fields, $q(\vb*x,t)$, by defining the spatially-integrated bispectrum,
\beq\label{eq:integral_bispectrum}
b(f_k,f_l) = \mathrm E\qty{ \int_\Omega \hat q^*(\vb*x,f_k)\hat q^*(\vb*x,f_l)\hat q(\vb*x,f_k+f_l) \dd{\vb*x}} = \mathrm E\{ \expval*{\hat{\vb*q}_{k+l},\hat{\vb*q}_{k\circ l}}_{\vb*x}\},
\eeq
where $\hat{\vb*q}$ are the spatially-resolved, temporal Fourier modes computed from the data, $\hat{\vb*q}_{k\circ l}\coloneq\hat{\vb*q}_k\circ \hat{\vb*q}_l$, $\circ$ denotes a Hadamard product, and $(\cdot)^*$ denotes the conjugate (transpose). The weighted inner product, $\expval*{\cdot,\cdot}_{\vb*x}$, is the same as that used for the SPOD in \S \ref{sec:energetics}.
Our goal is to measure the expected quadratic phase coupling between the components $\hat{\vb*q}_{k+l}$ and $\hat{\vb*q}_{k\circ l}$, independently of the power of each component. To this end, we normalise each component by the square root of its average integral power, $\sqrt{\mathrm E\{\norm*{\hat{\vb*q}_{k+l}}^2_{\vb*x}\}}$ and $\sqrt{\mathrm E\{\norm*{\hat{\vb*q}_{k\circ l}}^2_{\vb*x}\}}$, respectively, with the norm defined as $\norm{\vb*q}^2_{\vb*x}=\expval*{\vb*q,\vb*q}_{\vb*x}$. Thus normalised, the integrated bispectrum becomes equivalent to the integrated bicoherence,
\beq\label{eq:integral_bicoherence}
b(f_k,f_l) = \frac{\mathrm E\{ \expval*{\hat{\vb*q}_{k+l},\hat{\vb*q}_{k\circ l}}_{\vb*x}\}}{\sqrt{\mathrm E\{\norm*{\hat{\vb*q}_{k+l}}^2_{\vb*x}\} \mathrm E\{\norm*{\hat{\vb*q}_{k\circ l}}^2_{\vb*x}\}}}.
\eeq
Like the classical bicoherence for one-dimensional signals, the magnitude of the integrated bicoherence is bounded. Specifically,
\begin{subequations}
\begin{align}
|b(f_k,f_l)| &= \frac{|\mathrm E\{ \expval*{\hat{\vb*q}_{k+l},\hat{\vb*q}_{k\circ l}}_{\vb*x}\}|}{\sqrt{\mathrm E\{\norm*{\hat{\vb*q}_{k+l}}^2_{\vb*x}\} \mathrm E\{\norm*{\hat{\vb*q}_{k\circ l}}^2_{\vb*x}\}}} \\
&\leq \frac{\mathrm E\{| \expval*{\hat{\vb*q}_{k+l},\hat{\vb*q}_{k\circ l}}_{\vb*x}|\}}{\sqrt{\mathrm E\{\norm*{\hat{\vb*q}_{k+l}}^2_{\vb*x}\} \mathrm E\{\norm*{\hat{\vb*q}_{k\circ l}}^2_{\vb*x}\}}} \tag{triangle inequality} \\
&\leq \frac{\mathrm E\{ \norm*{\hat{\vb*q}_{k+l}}_{\vb*x}\norm*{\hat{\vb*q}_{k\circ l}}_{\vb*x}\}}{\sqrt{\mathrm E\{\norm*{\hat{\vb*q}_{k+l}}^2_{\vb*x}\} \mathrm E\{\norm*{\hat{\vb*q}_{k\circ l}}^2_{\vb*x}\}}} \leq1. \tag{Cauchy-Schwarz inequality}
\end{align}
\end{subequations}
In the following, we will assume the Fourier modes have been normalised to unit expected integral power.

For each triad, to estimate the BMD mode and mode bispectrum, $n_\mathrm{blk}$ independent realisations of the Fourier modes are assembled in the weighted data matrices
\begin{subequations}\label{eq:bmd_data_matrices}
\begin{align}
\hat{\mathsfbi Q}_{k\circ l} &= \frac{\mathsfbi W^{1/2} [ \hat{\vb*q}^{(1)}_{k\circ l}, \hat{\vb*q}^{(2)}_{k\circ l}, \cdots, \hat{\vb*q}^{(n_\mathrm{blk})}_{k\circ l} ]}{\norm{\mathsfbi W^{1/2} [ \hat{\vb*q}^{(1)}_{k\circ l}, \hat{\vb*q}^{(2)}_{k\circ l}, \cdots, \hat{\vb*q}^{(n_\mathrm{blk})}_{k\circ l} ]}_F} \\
\qand \hat{\mathsfbi Q}_{k+l} &= \frac{\mathsfbi W^{1/2} [ \hat{\vb*q}^{(1)}_{k+l}, \hat{\vb*q}^{(2)}_{k+l}, \cdots, \hat{\vb*q}^{(n_\mathrm{blk})}_{k+l} ]}{\norm{\mathsfbi W^{1/2} [ \hat{\vb*q}^{(1)}_{k+l}, \hat{\vb*q}^{(2)}_{k+l}, \cdots, \hat{\vb*q}^{(n_\mathrm{blk})}_{k+l} ]}_F},
\end{align}
\end{subequations}
which have been normalised such that $\|\hat{\mathsfbi Q}_{k\circ l}\|_F = \|\hat{\mathsfbi Q}_{k+l}\|_F = 1$, where $\norm{\cdot}_F$ is the Frobenius norm. In principle, a factor of $1/\sqrt{\nblk}$ is present in the numerator and denominator of each $\hat{\mathsfbi Q}$, which cancels out when $\hat{\mathsfbi Q}$ is normalised. For both $\hat{\mathsfbi Q}_{k\circ l}$ and $\hat{\mathsfbi Q}_{k+l}$, by definition,
\begin{equation}
\|\hat{\mathsfbi Q}\|_F = \sqrt{\mathrm{tr}(\hat{\mathsfbi Q}^*\hat{\mathsfbi Q})} = \frac{\sqrt{\sum_{i=1}^{\nblk} (\hat{\vb*q}^{(i)})^* \mathsfbi W \hat{\vb*q}^{(i)}}}{\norm{\mathsfbi W^{1/2} [ \hat{\vb*q}^{(1)}, \hat{\vb*q}^{(2)}, \cdots, \hat{\vb*q}^{(n_\mathrm{blk})} ]}_F},
\end{equation}
where $\mathrm{tr(\cdot)}$ denotes the trace. The Frobenius norm thus expresses the square root of the average power in $ \hat{\vb*q}$, which we normalise to one.
The weight and spectral estimation parameters are identical to those used for SPOD (see table~\ref{tab:SPOD_params}). We can define a single set of expansion coefficients, $\vb*a_{k,l}$, that relates the cross-frequency field, $\vb*\phi_{k\circ l}=\mathsfbi W^{-1/2}\hat{\mathsfbi Q}_{k\circ l}\vb*a_{k,l}$, and bispectral mode, $\vb*\phi_{k+l}=\mathsfbi W^{-1/2}\hat{\mathsfbi Q}_{k+l}\vb*a_{k,l}$, to the corresponding data matrices in equation~\eqref{eq:bmd_data_matrices}. BMD then seeks the particular set of expansion coefficients that maximises the magnitude of the mode bispectrum estimated from $\vb*\phi_{k\circ l}$ and $\vb*\phi_{k+l}$. That is,
\beq\label{eq:bmd_rayleigh}
\vb*a_{k,l} = \mathop{\arg\max}\limits_{\norm{\vb*a}=1} | \vb*\phi^*_{k\circ l}\mathsfbi W\vb*\phi_{k+l} | = \mathop{\arg\max}\limits_{\vb*a} \qty|\frac{\vb*a^*\mathsfbi B_{k,l}\vb*a}{\vb*a^*\vb*a}|,
\eeq
where $\mathsfbi B_{k,l} = \hat{\mathsfbi Q}^*_{k\circ l}\hat{\mathsfbi Q}_{k+l}$ is the bispectral density matrix. The mode bispectrum is recovered as $\beta_{k,l} = \vb*a_{k,l}^*\mathsfbi B_{k,l}\vb*a_{k,l}$, $\beta_{k,l}\in\mathbb C$. Equation~\eqref{eq:bmd_rayleigh} is the numerical radius problem for $\mathsfbi B_{k,l}$.

The magnitude of the BMD mode bispectrum is similarly bounded by unity, which can be proven as follows. The numerical radius of a matrix is bounded by its spectral norm \citep{GoldbergTadmor1982LAA,HornJohnson1985CUP},
\beq
|\beta(\mathsfbi B_{k,l})| \le \|\mathsfbi B_{k,l}\|_2,
\eeq
with the latter also equal to the leading singular value, $\sigma_1(\mathsfbi B_{k,l})$. To obtain an upper bound on $|\beta(\mathsfbi B_{k,l})|$, it thus suffices to consider $\|\mathsfbi B_{k,l}\|_2$. From the submultiplicativity of the spectral norm \citep{HornJohnson1985CUP}, it follows that
\beq
\|\mathsfbi B_{k,l}\|_2 = \|\hat{\mathsfbi Q}^*_{k\circ l}\hat{\mathsfbi Q}_{k+l}\|_2 \le \|\hat{\mathsfbi Q}^*_{k\circ l}\|_2 \|\hat{\mathsfbi Q}_{k+l}\|_2 = \|\hat{\mathsfbi Q}_{k\circ l}\|_2 \|\hat{\mathsfbi Q}_{k+l}\|_2.
\eeq
Since the spectral norm is bounded by the Frobenius norm \citep{HornJohnson1985CUP},
\beq
\|\hat{\mathsfbi Q}_{k\circ l}\|_2 \|\hat{\mathsfbi Q}_{k+l}\|_2 \le \|\hat{\mathsfbi Q}_{k\circ l}\|_F \|\hat{\mathsfbi Q}_{k+l}\|_F = 1.
\eeq
In other words, the mode bispectrum is bounded by the square root of the product of the average power in $\hat{\vb*q}_{k\circ l}$ and $\hat{\vb*q}_{k+l}$, in this case normalised to one.

%

In the original implementation of BMD by \citet{Schmidt2020NDyn}, the numerical radius is calculated using the iterative algorithm of \citet{HeWatson1997IMA}. The algorithm is not guaranteed to converge. In practice, non-convergence often leads to a mode bispectrum with a noisy appearance. In this work, we propose an improved implementation of BMD that utilises the iterative algorithm of \citet{MengiOverton2005IMA}, which is guaranteed to converge to the global maximum. The clean appearance of the mode bispectra later reported is a direct consequence of the robustness of this new implementation.

\begin{figure}
    \centering
    \includegraphics{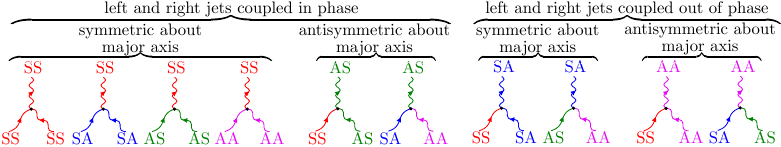}
    \caption{Non-redundant $D_2$ symmetry triads, colour-coded by symmetry. The mode bispectra of the SS-SS interaction, (SS,SS,SS), AS-AS interaction, (AS,AS,SS), and SS-AS interaction, (SS,AS,AS), are shown in figure \ref{fig:bmd_p_forced_spec_feynman}. The remaining triads are shown in figure \ref{fig:bmd_p_forced_spec_inactive}.}
    \label{fig:symmetry_triad_illustration}
\end{figure}

In addition to considering frequency triads $(f_k,f_l,f_k+f_l)$, we will also use BMD to investigate the nonlinear coupling among triadically-compatible $D_2$ symmetry components. Using classical bispectral analysis, \citet{WalkerThomas1997POF} previously uncovered experimental evidence for symmetry triads in supersonic single rectangular jets, with the caveat that symmetry was only assessed about one axis, which precludes the distinction between all four $D_2$ symmetry components. In the present work, $D_2$ symmetry triads are examined by assembling the Fourier modes $\hat{\vb*q}_k$, $\hat{\vb*q}_l$, and $\hat{\vb*q}_{k+l}$ from different, triadically compatible symmetry components. Ten of these non-redundant symmetry triads exist, and are illustrated in figure \ref{fig:symmetry_triad_illustration}. 

\subsubsection{Recovery of SPOD along the abscissa or ordinate}\label{sec:bmd_recover_spod}
\begin{figure}
    \centering
    \includegraphics{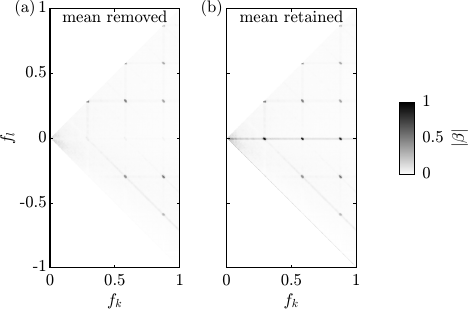}
    \caption{BMD mode bispectra of SS-SS interactions: (a) the long-time mean is removed from the data; (b) the long-time mean is included and SPOD is recovered on the $f_l=0$ axis (see appendix \ref{sec:appBmd_recover_spod}).}
    \label{fig:bmd_mean_inclusion}
\end{figure}
In the limit of infinitely long data and high frequency resolution, if the long-time mean is removed, the BMD mode bispectrum should vanish along $f_l=0$, $f_k=0$, and $f_{k+l}=0$. In practice, for limited data, the bispectrum tends to display finite values along these lines, indicative of unresolved low-frequency structures or slow trends captured by the zero-frequency bin of the DFT. Since the zero-frequency Fourier mode is real, it carries no phase information. For any triad that includes frequency zero, i.e., $(f,0,f)$, $(0,f,f)$, or $(f,-f,0)$, it can be shown that the triple correlation in the classical bispectrum defined in equation \eqref{eeq:classicBispectrum} may be rewritten as
\beq
\hat q^*(f)\hat q^*(0)\hat q(f) = \hat q^*(0)\hat q^*(f)\hat q(f) = \hat q^*(f)\hat q^*(-f)\hat q(0) = \hat q(0)|\hat q(f)|^2,
\eeq
which is a real quantity. Here, we invoked the conjugate symmetry of the Fourier transform of real data. The spatial integral of this triple correlation, equivalent to the inner product $\expval*{\hat{\vb*q}_{k+l},\hat{\vb*q}_{k\circ l}}_{\vb*x}$, is also real. The expectation, $\mathrm E\{ \expval*{\hat{\vb*q}_{k+l},\hat{\vb*q}_{k\circ l}}_{\vb*x}\}$, which defines the integral bispectrum in equation \eqref{eq:integral_bispectrum}, thus cannot measure the strength of phase coupling for such triads. Along these special lines, the classical and integral bispectra (and bicoherence) become phase-blind, and are sensitive only to a weighted (by the zero-frequency Fourier mode) version of power.


This observation motivates an extension of BMD to recover the SPOD spectrum and modes along the abscissa, $f_l=0$, or ordinate, $f_k=0$. For flow data with a spatially uniform mean, $\bar q$, recovery of the SPOD can be achieved simply by not subtracting the mean when solving the BMD problem. Neglecting block-to-block variations of the mean, which are likely small relative to $\bar q$, the zero-frequency Fourier mode corresponds to the constant scalar mean, $\bar q$. Along the abscissa, for $f_l=0$, the normalised bispectral density matrix simplifies to
\beq\label{eq:Bk,0}
\mathsfbi B_{k,0} = \hat{\mathsfbi Q}^*_{k\circ 0}\hat{\mathsfbi Q}_{k} = \hat{\mathsfbi Q}^*_k\hat{\mathsfbi Q}_k.
\eeq
Because $\mathsfbi B_{k,0}$ is now Hermitian and, by extension, normal, its numerical radius problem as given by equation~\eqref{eq:bmd_rayleigh} coincides with the method-of-snapshots SPOD eigenvalue problem in equation~\eqref{eq:spod_small_evp} \citep{GoldbergTadmor1982LAA,HornJohnson1985CUP}. Along $f_l=0$, BMD hence recovers the leading SPOD eigenvalues and modes of the normalised data. Similarly, along $f_k=0$,
\beq\label{eq:B0,l}
\mathsfbi B_{0,l} = \hat{\mathsfbi Q}^*_{0\circ l}\hat{\mathsfbi Q}_{l} = \hat{\mathsfbi Q}^*_l\hat{\mathsfbi Q}_l,
\eeq
which recovers the same SPOD spectrum and modes.

No SPOD is recovered along $f_{k+l}=0$, even with the uniform mean included. Again noting the conjugate symmetry of the Fourier transform of real data, the bispectral density matrix may be written as
\beq\label{eq:Bk,-k}
\mathsfbi B_{k,-k} = \hat{\mathsfbi Q}^*_{k\circ(-k)}\hat{\mathsfbi Q}_0 = (|\hat{\mathsfbi Q}_k|^2)^\mathrm{T}\hat{\mathsfbi Q}_0,
\eeq
where $|\cdot|^2$ denotes the element-wise squared absolute value.
This is the correlation between the mean and the squared magnitude of the Fourier mode $\hat{\vb*q}_k$, and is clearly not Hermitian, and thus does not correspond to an SPOD problem.

Whether SPOD is recovered on the abscissa or ordinate of BMD depends also on $D_2$ symmetry considerations. Because the mean flow has SS symmetry, SPOD can only be recovered from the symmetry triads that include SS as the first or second component. Specifically, among the non-redundant triads in figure \ref{fig:symmetry_triad_illustration}, the BMD for (SS,SS,SS), (SS,SA,SA), (SS,AS,AS), and (SS,AA,AA) recover the SPOD for SS, SA, AS, and AA, respectively. In the case of SS-SS interactions, both the abscissa and ordinate yield the SPOD. For SS-SA, SS-AS, and SS-AA interactions, due to the (arbitrary) ordering of the symmetries, the ordinate yields the SPOD.

For inhomogeneous flows, the turbulent mean is unlikely to be truly uniform. However, for the present jet data, the mean pressure is nearly uniform (see figure \ref{fig:mean_p_x}). Even close to the nozzle, at $x=5$, the mean pressure deviates locally from the ambient pressure, $p_\infty$, by at most four and five percent for the natural and forced jets, respectively. The relationships in equations \eqref{eq:Bk,0} and \eqref{eq:B0,l} are therefore satisfied approximately, as we demonstrate in detail in appendix \ref{sec:appBmd_recover_spod}. Figure~\ref{fig:bmd_mean_inclusion} compares the BMD mode bispectra for SS-SS interactions when the mean is removed or included. A full discussion of the bispectrum is deferred until \S \ref{sec:bisp_symm}. Here, we only remark that the bispectrum with the mean included shows higher magnitudes along $f_l=0$ and $f_{k+l}=0$ than the bispectrum without the mean. The mean-included bispectrum also displays local maxima along $f_l=0$, consistent with the forced SPOD spectrum for SS in figure \ref{fig:spod_p_spec}(a). The remainder of the bispectrum is independent of mean inclusion or removal. In what follows, we perform BMD exclusively with the mean included, combining spectral and bispectral information, that is, energetics and triadic coupling, in a single plot.

\subsection{Notes on symmetries of the mode bispectrum under \texorpdfstring{$D_2$}{D2} decomposition}\label{sec:bisp_symm}
Prior to interpreting the physics, we remark on some important symmetries of the mode bispectrum that the reader might observe. We note, however, that these bispectral symmetries are not salient in the physical interpretation. For discussions of the physics, the reader may safely skip to \S \ref{sec:bispec_analysis}.

\begin{figure}
    \centering
    \includegraphics{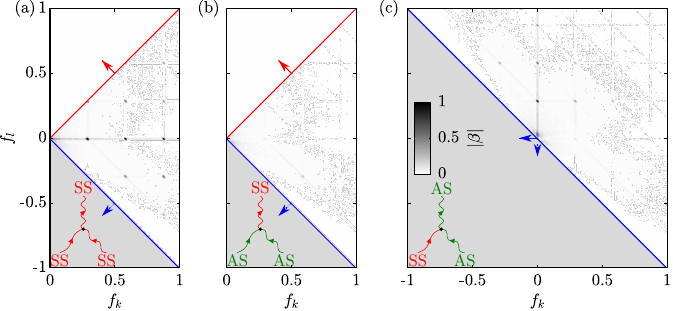}
    \caption{BMD mode bispectra: (a) SS-SS interactions; (b) AS-AS interactions; (c) SS-AS interactions. All three bispectra share the same contour levels. Each bispectrum displays only its non-redundant region. The redundant regions can be recovered from the non-redundant regions via reflection (solid arrow, ---$\!\blacktriangleright$) or reflection and complex conjugation (dotted arrow, $\cdots\!\blacktriangleright$).}
    \label{fig:bmd_p_forced_spec_feynman}
\end{figure}
Of the ten possible $D_2$ symmetry triads illustrated in figure \ref{fig:symmetry_triad_illustration}, only the (SS,SS,SS), (AS,AS,SS), and (SS,AS,AS) triads reveal clear signatures of active nonlinear interactions. The (SS,SS,SS) and (AS,AS,SS) triads involve interactions between coherent structures with the same symmetry, which we term symmetry-self interactions. Symmetry-self interactions, like AS-AS, yield a mode with SS symmetry. Conversely, the (SS,AS,AS) triad involves interactions between structures with different symmetries, which we term symmetry-cross interactions. The BMD mode bispectra of the SS-SS, AS-AS, and SS-AS interactions are shown in figure \ref{fig:bmd_p_forced_spec_feynman}(a,b,c), respectively. Only the non-redundant regions are shown \citep{Schmidt2020NDyn}. Each bispectrum displays a grid pattern of local maxima that indicate phase coupling between interacting waves. 
For the symmetry-self interactions in figure \ref{fig:bmd_p_forced_spec_feynman}(a,b), the mode bispectrum is symmetric about the red line, $f_l=f_k$, because the first two symmetry components can simply be swapped. The white region, $(f_k>0,f_l>f_k)$, can thus be recovered from a reflection of the non-redundant region about the red line, i.e.,
\beq
\beta(f_k,f_l)=\beta(f_l,f_k) \qfor f_k>0,~f_l>f_k.
\eeq
Owing to the conjugate symmetry of the Fourier transform of real data, i.e., $\hat{p}^*(f) = \hat{p}(-f)$, the mode bispectrum is also symmetric about the blue line, $f_l=-f_k$. The grey region, $(f_k>0,f_l<-f_k)$, can be recovered from a reflection of the non-redundant region about the blue line, followed by complex conjugation, i.e.,
\beq
\beta(f_k,f_l)=\beta^*(-f_l,-f_k) \qfor f_k>0,~f_l<-f_k.
\eeq
Similarly, the symmetry-cross interactions in figure \ref{fig:bmd_p_forced_spec_feynman}(c) enjoy conjugate symmetry. In this case, all of $f_l>-f_k$ is non-redundant. The grey region, $f_l<-f_k$, can be recovered from the former via a reflection about $f_k=0$, then another reflection about $f_l=0$, followed by conjugation. Concisely,
\beq
\beta(f_k,f_l)=\beta^*(-f_k,-f_l) \qfor f_l<-f_k.
\eeq
Implicit in figure \ref{fig:bmd_p_forced_spec_feynman}(c) is an additional symmetry between SS-AS and AS-SS interactions. It can be inferred from the form of the bispectral density that the mode bispectra of this pair of interactions are identical to each other under reflection about $f_l=f_k$. In other words,
\beq\label{eq:SSAS=ASSS}
\beta_{\text{SS-AS}}(f_k,f_l)=\beta_{\text{AS-SS}}(f_l,f_k).
\eeq
The subscript, $(\cdot)_\text{SS-AS}$, denotes the spatial symmetry corresponding to each frequency component: SS and AS symmetries for the $f_k$ and $f_l$ components, respectively. The mode bispectral symmetries illustrated here also generalise to other spatial symmetry triads of the twin jet.

No bispectral symmetry exists for the SS-AS interaction in figure \ref{fig:bmd_p_forced_spec_feynman}(c) about $f_l=f_k$. That is,
\beq\label{eq:SSAS_neq_SSAS}
\beta_{\text{SS-AS}}(f_k,f_l)\neq\beta_{\text{SS-AS}}(f_l,f_k).
\eeq
In the non-redundant region, the strength of the $(f_k,f_l,f_{k+l})$ triad for the SS-AS interaction is, on average, higher for $f_l>f_k$ than for $f_l<f_k$. Though this is not important to the present study, the likely explanation is that at the same harmonic frequency the SS component is more energetic than the AS component.
Given the identity~\eqref{eq:SSAS=ASSS}, the statistical asymmetry expressed in inequality~\eqref{eq:SSAS_neq_SSAS} can be restated as
\beq\label{eq:SSAS_neq_ASSS}
\beta_{\text{SS-AS}}(f_k,f_l)\neq\beta_{\text{AS-SS}}(f_k,f_l),
\eeq
Specifically, for $f_l>f_k$, the $(f_k,f_l,f_{k+l})$ triad for the SS-AS interaction is more strongly coupled than the same frequency triad but for the AS-SS interaction. For $f_l<f_k$, the reverse is true.

The relationship between the signs of $f_k$ and $f_l$ allows us to further classify triadic interactions as sum or difference interactions. Sum interactions are characterised by frequency doublets that satisfy $f_kf_l>0$. They couple primary waves at frequencies $f_k$ and $f_l$ to a secondary wave at a higher frequency, $f_{k+l}=f_k+f_l$ \citep{Phillips1960JFM,KimPowers1979IEEE}. Conversely, difference interactions, which satisfy $f_kf_l<0$, couple primary waves to a secondary wave at a lower frequency. A special case of difference interactions are mean flow deformations, which obey the condition that $f_k=-f_l$. Throughout this work, we will sometimes refer to a triad whose secondary wave contributes to the primary wave of another triad as the precursor or progenitor of the latter. When using nomenclature like precursor or progenitor, we refer only to this primary-secondary relationship between the two triads, and do not imply a causal relationship. 

\subsection{Bispectral analysis}\label{sec:bispec_analysis}
The (SS,SS,SS) triad in figure \ref{fig:bmd_p_forced_spec_feynman}(a) reveals both sum and difference interactions. The strongest nonlinear phase coupling identified is the difference interaction between SS modes at the fundamental and third harmonic frequencies, with a mode bispectrum magnitude of $|\beta_{\text{SS-SS}}(3f_0,-f_0)|=0.61$. Overall, however, the sum and difference interactions are of comparable coupling strength. On the other hand, the (AS,AS,SS) triad in figure \ref{fig:bmd_p_forced_spec_feynman}(b) is significantly biased towards difference interactions. No sum interaction is detected along $f_k=f_l$. The most strongly coupled frequency triad is the difference interaction between AS modes at the fundamental and second harmonic frequencies, with a magnitude of $|\beta_{\text{AS-AS}}(2f_0,-f_0)|=0.29$. The SS-AS mode bispectrum in figure \ref{fig:bmd_p_forced_spec_feynman}(c) shows strong sum interactions in the $(f_k>0,f_l>0)$ quadrant as well as strong difference interactions in the $(f_k<0,f_l>0)$ quadrant, but weak difference interactions in the $(f_k>0,f_l<0)$ quadrant. The strongest triad is the sum interaction between an SS second harmonic mode and an AS fundamental mode, with a magnitude of $|\beta_{\text{SS-AS}}(2f_0,f_0)|=0.37$. In a BMD mode bispectrum, sum interactions are an indicator of the classic frequency-doubling cascade that links the fundamental frequency component to successively higher harmonics. As figure \ref{fig:bmd_p_forced_spec_feynman} shows, both (SS,SS,SS) and (SS,AS,AS) symmetry triads participate in such cascades, whereas (AS,AS,SS) does not. This observation supports the hypothesis that while SS-SS and SS-AS interactions nonlinearly generate harmonics in SS and AS, respectively, AS-AS interactions do not generate harmonics in SS.

\begin{figure}
    \centering
    \includegraphics{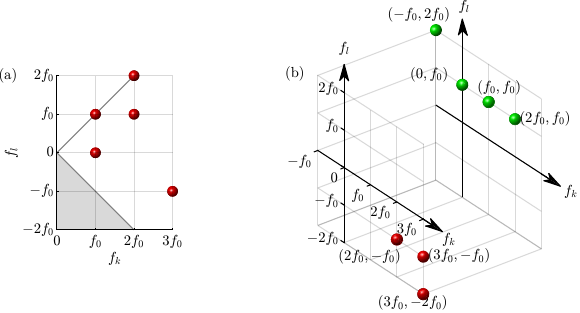}
    \caption{Dominant triads from figure \ref{fig:bmd_p_forced_spec_feynman}: (a) SS-SS interactions; (b) AS-AS and SS-AS interactions. The SS-SS and AS-AS interactions, which couple to modes with SS symmetry, are represented by red spheres. The SS-AS interactions, which couple to modes with AS symmetry, are represented by green spheres.}
    \label{fig:bmd_p_forced_ball_overview}
\end{figure}
In the following, we focus on symmetric mode couplings via the (SS,SS,SS) triad, and antisymmetric-symmetric mode couplings via the (AS,AS,SS) and (SS,AS,AS) triads. A representative set of dominant (SS,SS,SS) triads from figure \ref{fig:bmd_p_forced_spec_feynman}(a) are reproduced in figure \ref{fig:bmd_p_forced_ball_overview}(a), which we will interpret in the following. The triadic network is formed not from a limited set of triadic cascades, but should rather be understood as the collective interaction of all temporal and spatial scales identified by the BMD bispectrum and modes. As such, we will not exhaustively catalogue all possible pathways. Instead, in \S \ref{sec:SSself} we will focus on the most salient interactions that link the dominant triads in figure \ref{fig:bmd_p_forced_ball_overview}(a).

The (AS,AS,SS) triad cannot form a closed triad network, that is, the network cannot be visualised in a single bispectrum. This is because two primary waves with AS symmetry contribute to a secondary wave with SS symmetry. This secondary wave, however, cannot serve as the progenitor of another (AS,AS,SS) triad. Analogously, in an (SS,AS,AS) triad, primary waves with SS and AS symmetries contribute to a secondary wave with AS symmetry. This secondary wave cannot provide the SS component of another (SS,AS,AS) triad, so the network is again unclosed. The combination of the two symmetry triads, on the other hand, does form a closed network. SS-SS and SS-AS interactions contribute to secondary waves with both AS and SS symmetries. The secondary AS waves may in turn serve as the primary waves in another (AS,AS,SS) triad. Similarly, a pair of secondary AS and SS waves may participate in another (SS,AS,AS) triad. To elucidate these inter-triad relationships, we combine the dominant triads from the AS-AS and SS-AS bispectra in figure \ref{fig:bmd_p_forced_spec_feynman}(b,c) into a three-dimensional bispectrum, shown in figure \ref{fig:bmd_p_forced_ball_overview}(b). The dominant (AS,AS,SS) triads are displayed on one pair of frequency axes, the dominant (SS,AS,AS) triads on another. The pathways between them will be examined in detail in \S \ref{sec:ASself_SSAS}.

\begin{figure}
    \centering
    \includegraphics{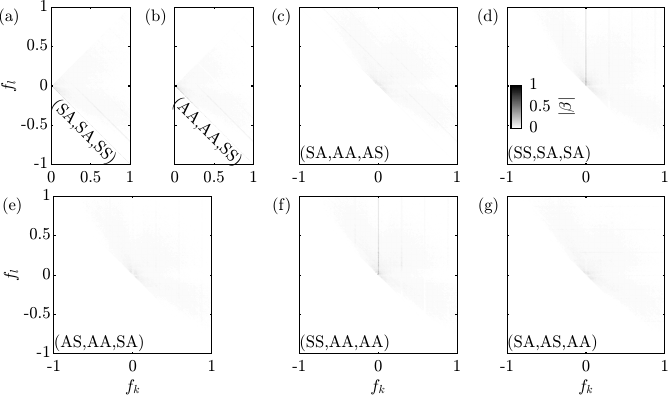}
    \caption{BMD mode bispectra of the symmetry triads not shown in figure \ref{fig:bmd_p_forced_spec_feynman}. All panels use the same contour levels as figure \ref{fig:bmd_p_forced_spec_feynman}.}
    \label{fig:bmd_p_forced_spec_inactive}
\end{figure}
The seven symmetry triads not shown in figure \ref{fig:bmd_p_forced_spec_feynman} are reported in figure \ref{fig:bmd_p_forced_spec_inactive}. These mode bispectra lack the telltale grid pattern of local maxima that would have otherwise been indicative of triadic interactions. With the exception of the $f_k=0$ lines in figure \ref{fig:bmd_p_forced_spec_inactive}(d) and \ref{fig:bmd_p_forced_spec_inactive}(f), which recover the leading SPOD spectra of SA and AA, respectively, the magnitudes of all seven bispectra are very small. All of these triads include either SA or AA symmetry (or both) as one or more components. We conclude from this that the SA and AA symmetries do not take part in nonlinear interactions involving the forcing frequency.

\subsection{Symmetric mode interactions}\label{sec:SSself}
\begin{figure}
    \centering
    \includegraphics{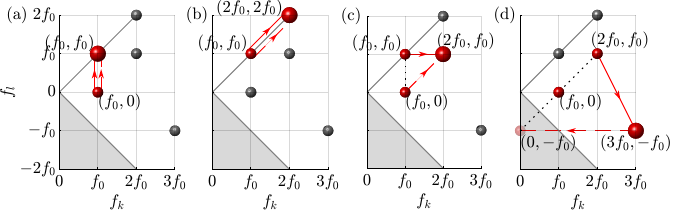}
    \caption{Five representative triads from the (SS,SS,SS) mode bispectrum in figure \ref{fig:bmd_p_forced_spec_feynman}(a). Large red spheres highlight the following triads: (a) $(f_0,f_0,2f_0)_\text{SS-SS-SS}$; (b) $(2f_0,2f_0,4f_0)_\text{SS-SS-SS}$; (c) $(2f_0,f_0,3f_0)_\text{SS-SS-SS}$; (d) $(3f_0,-f_0,2f_0)_\text{SS-SS-SS}$. Their possible precursor triads are marked by the small red spheres. All panels share the same axes. In each panel, the large sphere indicates the secondary wave at $f_{k+l}$. Red solid (\textcolor{red}{---}) and dashed (\textcolor{red}{-- --}) lines distinguish between $f_k$ and $f_l$, respectively. Dotted lines ($\cdots$) connect $f_k$ to $f_l$. Arrows point towards or away from $f_{k+l}$ in sum or difference interactions, respectively. The translucent sphere denotes a complex conjugate.}
    \label{fig:bmd_p_forced_spec2d_SS}
\end{figure}
\begin{figure}
    \centering
    \includegraphics{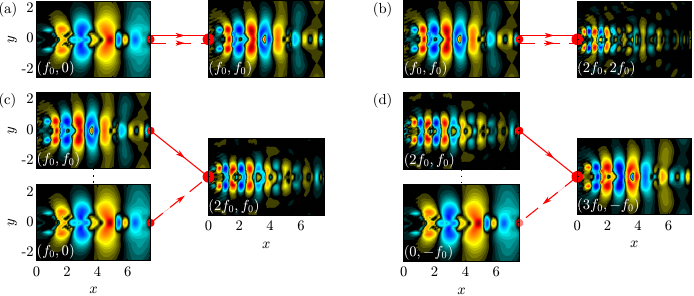}
    \caption{Bispectral modes of the SS-SS interactions in figure \ref{fig:bmd_p_forced_spec2d_SS}. The corresponding interactions and modes are labelled with the same panel indices, (a--d). The $z=1.8$ plane is displayed. In this and the following figures of BMD modes, the colours saturate at $|\vb*\phi|/\max|\vb*\phi|=\pm1$. See supplementary movie 1 for an animation.}
    \label{fig:bmd_p_forced_mode_SS}
\end{figure}
The grid of local maxima in figure \ref{fig:bmd_p_forced_spec_feynman}(a) implies an intricate network of interconnected symmetry-self interactions between modes that possess SS symmetry. In figure \ref{fig:bmd_p_forced_spec2d_SS} we illustrate them using five representative triads. The corresponding bispectral modes are shown in figure \ref{fig:bmd_p_forced_mode_SS}. For brevity, we display only the $z=1.8$ plane, which passes through the centreline of one nozzle and is parallel to the minor-axis plane. This choice is motivated by the observation that the SS and AS spatial symmetries, which are the only two components to engage in triadic interactions, are both symmetric about the minor-axis plane. They are thus most easily distinguished by their opposing symmetries about the major-axis plane, $y=0$. 

Figure~\ref{fig:bmd_p_forced_spec2d_SS}(a) considers the $(f_0,f_0,2f_0)_\text{SS-SS-SS}$ triad, where a mode at the fundamental forcing frequency, $f_0$, is coupled to a mode at the second harmonic, $2f_0$. To each primary-wave frequency, $f_{k}$ or $f_{l}$, a number of triads can contribute collectively, in particular those indicated by the local maxima in the bispectrum along the diagonal of slope $-1$ associated with that $f_{k}$ or $f_{l}$. In this case, the first harmonic primary mode of the $(f_0,f_0,2f_0)_\text{SS-SS-SS}$ triad coincides with the secondary modes of a number of other triads, e.g., $(f_0,0,f_0)_\text{SS-SS-SS}$ and $(2f_0,-f_0,f_0)_\text{SS-SS-SS}$. Here we highlight the pathway from $(f_0,0,f_0)_\text{SS-SS-SS}$ to $(f_0,f_0,2f_0)_\text{SS-SS-SS}$. The $(f_0,0,f_0)_\text{SS-SS-SS}$ triplet recovers the leading SPOD mode and modal energy at $f_0$. As shown in figure \ref{fig:bmd_p_forced_mode_SS}(a), excitations delivered by the plasma actuators to the initial shear layer are spatially amplified by the well-known Kelvin-Helmholtz (KH) instability. The bispectral mode captures symmetric KH-type wavepackets with extended spatial support in the streamwise direction.
The mode shape bears qualitative resemblance to the symmetric modes extracted in the experiments of \citet{SamimyEtAl2023JFM} from the same twin-rectangular jet, albeit overexpanded and forced in a different pattern. Trapped in the potential core are waves of a higher wavenumber. We have confirmed that these core modes are also present in the SA component. Due to the low supersonic jet velocity, the core modes are equivalent to the subsonic instability waves identified and extensively investigated by \citet{TamHu1989JFM} and \citet{TowneEtAl2017JFM} in round jets. As they are transversely confined to the supersonic jet core in both the major- and minor-axis planes, the subsonic instability waves cannot propagate upstream (see supplementary movie 1 for an animation). Their maximum amplitude is found much closer to the nozzle than similar waves in supersonic, nearly-shock-free round jets \citep{SchmidtEtAl2018JFM,TowneEtAl2019AIAA}. For shock-containing jets, these waves are documented in detail by \citet{Edgington-MitchellEtAl2021JFM}. Evidence of these modes has previously surfaced in single-rectangular \citep{ZamanEtAl2015JFM,GojonEtAl2016JFM,GojonEtAl2019AIAAJ,SemlitschEtAl2020JFM,TamChandramouli2020JSV,ZamanEtAl2022JFM,FerreiraEtAl2023AIAAJ,WuEtAl2023arXiv,KarnamEtAl2023POF} as well as twin-rectangular jets \citep{SamimyEtAl2023JFM,JeunEtAl2024JFM}. Here, the trapped modes attain an amplitude comparable to the KH-type instability waves, suggesting they too are indirectly energised by the symmetric forcing.

The fundamental forcing mode, $(f_0,0,f_0)_\text{SS-SS-SS}$, is coupled to the second harmonic mode via the self interaction $(f_0,f_0,2f_0)_\text{SS-SS-SS}$. The second harmonic mode, at twice the frequency of the fundamental, has approximately double the dominant streamwise wavenumber as well. This follows from the linear dispersion relation of KH-type waves in the initial shear-layer region of jets \citep{SchmidtEtAl2018JFM}, and directly demonstrates that symmetry and frequency triads, which are enforced by our analysis, naturally form triads in wavenumber space. The wavepacket of the second harmonic mode is more compact, with its spatial support confined both axially and transversely, relative to the fundamental mode. Trapped modes are again observed. Compared to the KH-type waves, which are confined to $x\lesssim5$, the trapped waves appear to have more extended support in the axial direction. The presence of trapped waves in the bispectral mode $(f_0,f_0,2f_0)_\text{SS-SS-SS}$ suggests these waves are also quadratically phase-coupled with the fundamental mode. 

In figures~\ref{fig:bmd_p_forced_spec2d_SS}(b) and \ref{fig:bmd_p_forced_mode_SS}(b), the bispectral mode $(2f_0,2f_0,4f_0)_\text{SS-SS-SS}$ couples the second harmonic mode at frequency $2f_0$ to the fourth harmonic mode at frequency $4f_0$ through another frequency-doubling self interaction. Analogous to the relationship between the waveforms of the second harmonic and fundamental modes, the fourth harmonic mode also doubles the dominant streamwise wavenumber of the second harmonic mode. At the same time, the streamwise extent of the former is approximately halved, since only a thinner shear layer can support the higher-wavenumber KH-type wave. 
Similar trends are observed for the $(2f_0,f_0,3f_0)_\text{SS-SS-SS}$ triad in figures~\ref{fig:bmd_p_forced_spec2d_SS}(c) and \ref{fig:bmd_p_forced_mode_SS}(c), where the second harmonic and fundamental modes undergo sum interaction, thereby becoming coupled to the third harmonic mode at frequency $3f_0$. Together, the $(f_0,0,f_0)_\text{SS-SS-SS}$, $(f_0,f_0,2f_0)_\text{SS-SS-SS}$, $(2f_0,f_0,3f_0)_\text{SS-SS-SS}$, and $(2f_0,2f_0,4f_0)_\text{SS-SS-SS}$ triads form a hierarchy of coherent structures at four different frequencies, and exemplify a cascade of successively higher frequency and wavenumber components.

The $(3f_0,-f_0,2f_0)_\text{SS-SS-SS}$ triad in figure \ref{fig:bmd_p_forced_spec2d_SS}(d) indicates a difference interaction between the third harmonic and fundamental modes. One candidate for such an interaction is the coupling between the $(2f_0,f_0,3f_0)_\text{SS-SS-SS}$ and $(0,-f_0,-f_0)_\text{SS-SS-SS}$ bispectral modes. The latter is the complex conjugate of the $(f_0,0,f_0)_\text{SS-SS-SS}$ mode in the non-redundant region of the mode bispectrum. These three interacting modes are reported in figure \ref{fig:bmd_p_forced_mode_SS}(d). They illustrate the triadic coupling of a primary wave with high frequency and wavenumber, $(2f_0,f_0,3f_0)_\text{SS-SS-SS}$, to a secondary wave with lower frequency and wavenumber, $(3f_0,-f_0,2f_0)_\text{SS-SS-SS}$.

\subsection{Antisymmetric-symmetric mode interactions}\label{sec:ASself_SSAS}
\begin{figure}
    \centering
    \includegraphics{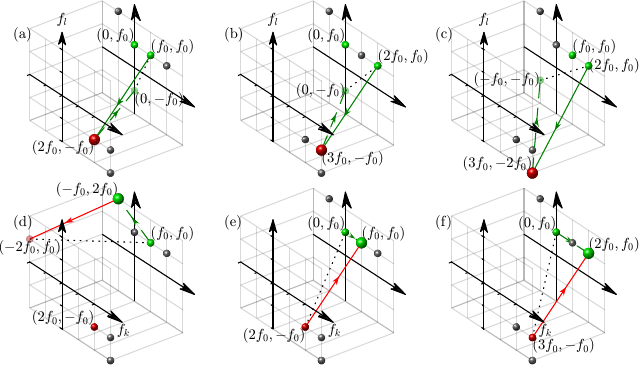}
    \caption{Same as figure \ref{fig:bmd_p_forced_spec2d_SS} but for the (AS,AS,SS) (red) and (SS,AS,AS) (green) symmetry triads. The following frequency triads are highlighted by large spheres: (a) $(2f_0,-f_0,f_0)_\text{AS-AS-SS}$; (b) $(3f_0,-f_0,2f_0)_\text{AS-AS-SS}$; (c) $(3f_0,-2f_0,f_0)_\text{AS-AS-SS}$; (d) $(-f_0,2f_0,f_0)_\text{SS-AS-AS}$; (e) $(f_0,f_0,2f_0)_\text{SS-AS-AS}$; (f) $(2f_0,f_0,3f_0)_\text{SS-AS-AS}$.  Their possible precursor triads are exemplified by small red or green spheres. As in figure \ref{fig:bmd_p_forced_spec2d_SS}, solid (---) and dashed (-- --) lines distinguish between $f_k$ and $f_l$, respectively, while dotted lines ($\cdots$) connect $f_k$ to $f_l$. Red (\textcolor{red}{$\bullet$}) and green (\textcolor{Green}{$\bullet$}) spheres represent SS and AS symmetries, respectively.}
    \label{fig:bmd_p_forced_spec3d}
\end{figure}
\begin{figure}
    \centering
    \includegraphics{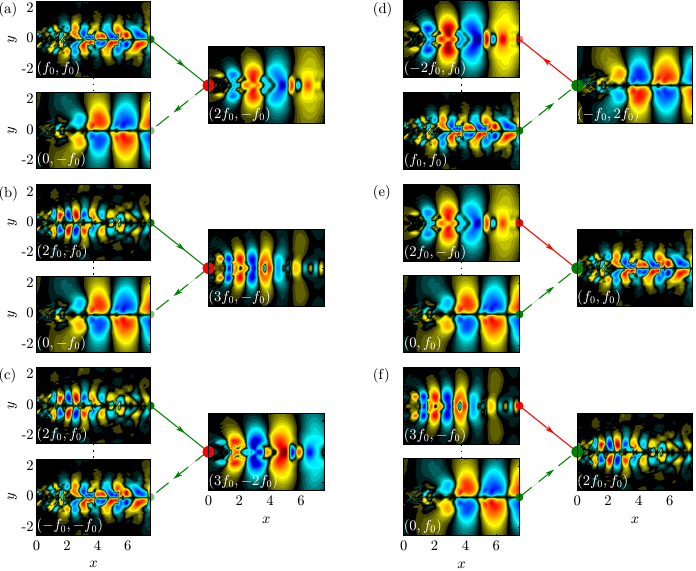}
    \caption{Same as figure \ref{fig:bmd_p_forced_mode_SS} but for AS-AS (left column) and SS-AS (right column) interactions. The panel indices, (a--f), correspond to the interactions in figure \ref{fig:bmd_p_forced_spec3d}. See supplementary movie 2 for an animation.}
    \label{fig:bmd_p_forced_mode_SS_AS}
\end{figure}
In this section, we examine the network of spatial symmetry triads that is enabled by the wave coupling between modal structures with SS and AS symmetries. As in \S \ref{sec:SSself}, we make no attempt to enumerate all such couplings. Rather, we focus on a truncated triad network made up of the most dominant triads, previously shown in figure \ref{fig:bmd_p_forced_ball_overview}(b), and reproduced in figure \ref{fig:bmd_p_forced_spec3d}. For each of these dominant triads, we investigate one pathway that could contribute to the triad.

As we observed in \S \ref{sec:bispec_analysis}, the (AS,AS,SS) symmetry triad mainly undergoes difference interactions. Figure~\ref{fig:bmd_p_forced_spec3d}(a) considers the coupling between an AS second harmonic mode, at frequency $2f_0$, an AS fundamental mode, at frequency $-f_0$, and an SS first harmonic mode, at frequency $f_0$. This coupling is detected in the AS-AS mode bispectrum as the $(2f_0,-f_0,f_0)_\text{AS-AS-SS}$ frequency triad. One of its possible precursors is the SS-AS bispectral mode of the $(0,-f_0,-f_0)_\text{SS-AS-AS}$ triad, shown in figure \ref{fig:bmd_p_forced_mode_SS_AS}(a) on the bottom left, which is the complex conjugate of the $(0,f_0,f_0)_\text{SS-AS-AS}$ triad, i.e., the leading AS SPOD mode at $f_0$. The $(0,-f_0,-f_0)_\text{SS-AS-AS}$ triad has a secondary wave frequency of $f_{k+l}=-f_0$, matching the primary wave frequency of the $(2f_0,-f_0,f_0)_\text{AS-AS-SS}$ triad, $f_l=-f_0$. The $(0,-f_0,-f_0)_\text{SS-AS-AS}$ bispectral mode is dominated by a wavepacket structure located in the shear layer. The wavepacket flaps antisymmetrically about the major axis plane, $y=0$. The envelope of the packet forms a standing wave in the near jet (see supplementary movie 2 for an animation), and is produced by the interference between downstream-propagating KH-type waves and upstream-propagating guided-jet modes (G-JM) or freestream acoustic waves \citep{Edgington-MitchellEtAl2022JFM}. This behaviour is well-known in screeching jets. It was first noted in the experiments by \citet{DaviesOldfield1962Acustica}, \citet{WestleyWoolley1968Toronto}, and others, and investigated in depth by \citet{Panda1999JFM}; see also \citet{Edgington-Mitchell2019IJA} for a recent review. 
No standing waves are visible in SS or SA bispectral modes, which are symmetric about the major axis. Examples include the SS first and second harmonic modes on the right of figures \ref{fig:bmd_p_forced_mode_SS_AS}(a) and \ref{fig:bmd_p_forced_mode_SS_AS}(b), respectively. The absence of symmetric standing waves in twin-rectangular jets is consistent with previous observations \citep{JeunEtAl2022AIAAJ,SamimyEtAl2023JFM}. 

The role of the complex conjugate in facilitating AS-AS difference interactions is also demonstrated by the $(3f_0,-f_0,2f_0)_\text{AS-AS-SS}$ triad in figure~\ref{fig:bmd_p_forced_spec3d}(b), with its corresponding modes in figure \ref{fig:bmd_p_forced_mode_SS_AS}(b). It couples a pair of AS modes at the third and first harmonics, with frequencies $3f_0$ and $-f_0$, respectively, to an SS mode at the second harmonic. One such example of a pair of AS primary modes consists of the SS-AS bispectral modes of the $(2f_0,f_0,3f_0)_\text{SS-AS-AS}$ and $(0,-f_0,-f_0)_\text{SS-AS-AS}$ triads. The latter is the conjugate of the $(0,f_0,f_0)_\text{SS-AS-AS}$ triad. Similarly, the $(3f_0,-2f_0,f_0)_\text{AS-AS-SS}$ triad in figure~\ref{fig:bmd_p_forced_spec3d}(c) couples third and second harmonic AS modes, with frequencies $3f_0$ and $-2f_0$, respectively, to an SS mode at the first harmonic. An example pair of primary waves contributing to the $(3f_0,-2f_0,f_0)_\text{AS-AS-SS}$ triad are the SS-AS bispectral modes resulting from the $(2f_0,f_0,3f_0)_\text{SS-AS-AS}$ and $(-f_0,-f_0,-2f_0)_\text{SS-AS-AS}$ triads. The latter is the conjugate of the $(f_0,f_0,2f_0)_\text{SS-AS-AS}$ triad. The corresponding modes are shown in figure \ref{fig:bmd_p_forced_mode_SS_AS}(c).

Whereas the triads reported in figure \ref{fig:bmd_p_forced_spec3d}(a--c) involve symmetry-self interactions between AS mode pairs, the triads in figure \ref{fig:bmd_p_forced_spec3d}(d--f) entail symmetry-cross interactions between SS and AS modes. Unlike AS-AS interactions, the most dominant SS-AS interactions include both sum and difference interactions. The $(-f_0,2f_0,f_0)_\text{SS-AS-AS}$ triad in figure \ref{fig:bmd_p_forced_spec3d}(d) indicates a difference interaction between an SS first harmonic mode at frequency $-f_0$ and an AS second harmonic mode at frequency $2f_0$. These primary waves couple to an AS secondary wave at the first harmonic, $f_0$. The AS bispectral mode of this triad is displayed on the right of figure \ref{fig:bmd_p_forced_mode_SS_AS}(d), and shows the KH and G-JM interference pattern discussed previously. The $(f_0,f_0,2f_0)_\text{SS-AS-AS}$ triad in figure \ref{fig:bmd_p_forced_spec3d}(e) indicates the sum interaction between two first harmonic modes with SS and AS symmetries and an AS secondary mode at the second harmonic. The bispectral mode at the second harmonic is shown on the right of figure \ref{fig:bmd_p_forced_mode_SS_AS}(e). It reveals both KH-type instability waves as well as core modes, but not clear signatures of G-JM. The AS core modes at the second harmonic share similar streamwise distribution as the SS core modes at the same frequency on the right of figure \ref{fig:bmd_p_forced_mode_SS_AS}(b), but differ in their distribution in $y$. Specifically, the AS core modes straddle the centreline, and have double the number of anti-nodes in $y$, as required by antisymmetry, compared to the SS core modes. In short, AS bispectral modes support both types of subsonic instability waves: those trapped in the core, and the G-JM that propagates in the shear layer and the ambient flow, with the preponderance of each type varying according to frequency. By contrast, SS bispectral modes appear to support only core modes. Sensitivity of the G-JM to $D_2$ symmetry has also been predicted using linear stability models for twin-round jets \citep{StavropoulosEtAl2023JFM}, and generalizes from circular symmetry for single-round jets \citep{TamHu1989JFM}. A subsonic jet mode can propagate upstream against the supersonic jet flow only if it has partial support outside the jet, in the slow ambient flow \citep{TamHu1989JFM}. Because of this, the difference in transverse spatial decay between subsonic modes with AS and SS symmetries, and between such modes at disparate frequencies, impacts the possibility and strength of feedback. This suggests an intimate link between the $D_2$ symmetry of the subsonic mode and its permitted direction of propagation. It forms part of the explanation for the symmetry-dependence of tones in the natural and forced twin-rectangular jets.

Analogous to the cascade of triads exemplified by the SS-SS interactions in figures~\ref{fig:bmd_p_forced_spec2d_SS} and \ref{fig:bmd_p_forced_mode_SS}, SS-AS interactions also facilitate the coupling between harmonic frequencies. In figure \ref{fig:bmd_p_forced_spec3d}(e), the SS-AS mode bispectrum detects the phase coupling between SS and AS modes at the first harmonic, and an AS mode at the second harmonic, enabled by the $(f_0,f_0,2f_0)_\text{SS-AS-AS}$ triad. Similarly, figure \ref{fig:bmd_p_forced_spec3d}(f) shows the $(2f_0,f_0,3f_0)_\text{SS-AS-AS}$ triad, which couples SS and AS modes at the second and first harmonics, respectively, to an AS mode at the third harmonic.
The corresponding bispectral modes are reported in figures~\ref{fig:bmd_p_forced_mode_SS_AS}(e) and \ref{fig:bmd_p_forced_mode_SS_AS}(f), respectively. They show the familiar doubling and tripling of axial wavenumbers expected from modes that span three successive frequency harmonics.

The AS-AS bispectral mode of the $(3f_0,-f_0,2f_0)_\text{AS-AS-SS}$ triad, shown on the right of figure \ref{fig:bmd_p_forced_mode_SS_AS}(b), is indistinguishable from the SS-SS mode belonging to the same frequency triad on the right of figure \ref{fig:bmd_p_forced_mode_SS}(d). The equivalence between the SS-SS and AS-AS bispectral modes indicates that the symmetric and antisymmetric structures at frequencies $3f_0$ and $-f_0$ are phase-locked to the same symmetric structure at $2f_0$. 
Although we have investigated the SS-SS bispectrum independently of the AS-AS and SS-AS bispectra, the close agreement between the SS-SS and AS-AS bispectral modes at the same $f_{k+l}$ confirms the notion that all three symmetry triads are in fact part of the same interconnected network of triad interactions.

\section{Discussion}\label{sec:discussion}
In this paper, we have extended BMD to the $D_2$ symmetry of the twin jet, and confirmed the presence of triads associated with the triple correlation between $D_2$ symmetry components. The detection of frequency and symmetry triads rests upon the three-wave resonance condition that must be fulfilled by both. Compatibility between frequency and symmetry components is duly enforced in BMD. It is perhaps less obvious that the same wave components also form triads in wavenumber space---an empirical finding not explicitly guaranteed by the algorithm, yet clearly borne out by the modes in figures~\ref{fig:bmd_p_forced_mode_SS} and \ref{fig:bmd_p_forced_mode_SS_AS}.

Recovering the leading SPOD eigenvalues and modes from BMD numerical radii and modes requires a sufficiently uniform mean flow. In appendix \ref{sec:appBmd_recover_spod}, figure \ref{fig:bmd_recover_spod}, we have confirmed that for the pressure 2-norm used throughout this study, the BMD bispectrum for SS-SS interactions along $f_l=0$ is identical to the leading SPOD spectrum for SS. We have also confirmed that this holds true for other observables, including density and temperature, whose mean is approximately uniform. If this condition is not satisfied, the direct equivalence between BMD and SPOD breaks down. In the latter case, it may nevertheless be desirable to recover the SPOD from the BMD. We achieve this by explicitly setting the zero-frequency component of the Fourier transform to unity, guaranteeing perfect recovery of the SPOD (see figure \ref{fig:bmd_unitmean_recover_spod}). 

BMD is not without shortcomings. It enables us to systematically catalogue quadratic phase coupling, which is a known mechanism for scale-to-scale energy transfer, and reveals the corresponding flow structures, but not to quantify the amount of energy transferred---a concern also voiced by \citet{FreemanEtAl2024JFM}. However, the analysis of energy transfer requires taking the exact form of the nonlinearity in the governing equations into account, and is often not possible when the full flow state is inaccessible, e.g. from schlieren measurements. 

Each of the BMD modes in figure \ref{fig:bmd_p_forced_mode_SS} for the (SS,SS,SS) triad and figure \ref{fig:bmd_p_forced_mode_SS_AS} for the (AS,AS,SS) and (SS,AS,AS) triads consists of two or more types of waves with distinct transverse spatial support (jet core, shear layer, or freestream) and directions of propagation (upstream or downstream). While these waves may each be explainable by a different physical interpretation (KH, core mode, or G-JM), they are coupled within a single coherent structure, and thus may be best understood as inextricable components of a single modal structure, rather than as distinct flow phenomena.

Some fundamental questions about the forced twin jet remain unanswered. The character of the tone at $f_0$ in the AS component (see figure \ref{fig:spod_p_spec}) is ambiguous. Our analysis does not allow us to categorise this spectral peak unequivocally as a screech tone or simply a response to the forcing. Since screech is suppressed in the AA component of the forced jet, it is conceivable that screech is also absent from AS. Nonetheless, the presence of higher harmonics in the AS component lends support to the notion that the $f_0$ tone is indeed a screech mode. Because only the SS component is forced, and SS-SS interactions only generate SS secondary waves, the AS higher harmonics must have arisen nonlinearly. This is consistent with the SS-AS bispectrum in figure \ref{fig:bmd_p_forced_spec_feynman}(c), which confirms that (SS,AS,AS) triads are highly active in the flow. Hypothetically, in the absence of AS screech, the strong SS forcing could interact with the underlying turbulence in the AS component and give birth to tones in AS. However, this fails to explain why tones with SA or AA symmetries are not present in the forced jet. In addition, high-amplitude, axisymmetric forcing of round turbulent jets has produced no evidence of the formation of non-axisymmetric tones \citep{NekkantiEtAl2023AIAA,HeidtColonius2024JFM}. More likely, the AS screech mode remains active in the forced jet as a linear stability mode. Its nonlinear interaction with the SS forcing then generates AS higher harmonics. Screech-forcing interactions in twin-rectangular jet flows were previously proposed by \citet{SamimyEtAl2023JFM} from experimental observations.

Future studies that force the twin-rectangular jet in the SA, AS, or AA components could provide clues to these questions. Forcing the antisymmetric components would also help establish (or refute) the generality of the phenomenon that dominant instabilities adopt only two symmetries at a time, e.g., AS and AA in the natural jet, SS and AS in the symmetrically-forced jet. Due to computational limitations, this is out of the scope of the present study.

To aid our investigation of the nonlinear dynamics of the forced jet, we introduce two innovations to BMD. First, consistent with the bicoherence measure for one-dimensional signals, we normalise BMD by the power of each frequency component such that the magnitude of the BMD mode bispectrum is bounded by unity. This enables the strength of nonlinear mode coupling to be interpreted directly and intuitively. Second, we show that under certain conditions, inclusion of the mean flow permits recovery of the leading SPOD eigenvalues and modes from BMD. Both features are added to our \textsc{Matlab} implementation of BMD, available at \url{https://www.mathworks.com/matlabcentral/fileexchange/83408-bispectral-mode-decomposition}.

\section{Summary and conclusions}\label{sec:conclusion}
We investigate the dominant physical instabilities and nonlinear dynamics of a supersonic twin-rectangular jet. LES of the jet are carried out, both in its natural state and forced using plasma actuation. In the natural jet, SPOD reveals screech tones in the AS and AA $D_2$-symmetry components. The modal structures associated with the AS and AA symmetries are antisymmetric about the major axis, i.e., they are flapping instabilities. The left and right jets are coupled either perfectly in phase (AS) or perfectly out of phase (AA). Time-frequency analysis based on SPOD expansion coefficients indicates the AS screech mode steadily dominates over all time, while the weaker AA screech mode is intermittent.

Given the antisymmetry of the screech modes, we test the hypothesis that screech can be mitigated by forcing the SS component of the twin jet. Upon the application of SS forcing at the natural screech frequency, the jet exhibits tones only in the SS and AS components, in which the dominant instabilities are symmetric (SS) or antisymmetric (AS) about the major axis, with the left and right jets coupled in phase. These tones are located at the fundamental forcing frequency and its harmonics. In the AS component, the energy at low and near-zero frequencies is significantly elevated. AA screech tones are entirely eliminated, whereas the SA component is unmodified by the SS forcing.

Applying BMD to the forced jet, we confirm the existence of triads within one symmetry component, as well as across different---but compatible---symmetries. Out of ten possible symmetry triads, only three are statistically significant: (SS,SS,SS), (AS,AS,SS), and (SS,AS,AS). Together, the SS and AS symmetries form an interconnected web of triad cascades. In the frequency space, (SS,SS,SS) and (SS,AS,AS) triads are characterised by both sum and difference interactions, leading to inter-frequency coupling among modes at a hierarchy of harmonic frequencies. (AS,AS,SS) interactions, on the other hand, are dominated by difference interactions.

The coherent structures educed using BMD highlight the primacy of two physical mechanisms in the SS and AS components: KH-type shear-layer instabilities, and the subsonic instability waves of \citet{TamHu1989JFM}. Both mechanisms are active in both symmetries. In the SS bispectral modes, the subsonic waves are trapped in the supersonic core region, and thus downstream-travelling. In the AS modes, these waves are also found in the shear layer in addition to the potential core. The shear-layer subsonic waves are the upstream-travelling G-JM implicated in screech resonance \citep{Edgington-MitchellEtAl2022JFM}. The distinct spatial support of AS subsonic waves explains why the twin-rectangular jet manifests screech modes with AS but not SS symmetry. The close association between flow symmetry and the existence of the G-JM is thus a general phenomenon that translates from the azimuthal symmetry of round jets to the dihedral group symmetry of twin-rectangular jets, and likely generalises to other types of statistical symmetries.



\backsection[Supplementary data]{\label{SupMat}Supplementary material and movies are available at \\https://doi.org/10.1017/jfm.2019...}

\backsection[Acknowledgements]{We thank M. Samimy, N. Webb, A. Esfahani, and R. Leahy for providing voltage and current measurements of their plasma actuators. O.T.S. thanks Julian Domaradzki for insightful discussions on triadic interactions.}

\backsection[Funding]{We gratefully acknowledge support from Office of Naval Research grant N00014-23-1-2457, under the supervision of Dr. Steve Martens. LES calculations were carried out on computational resources provided by DoD HPCMP at the ERDC DSRC supercomputer facility.}

\backsection[Declaration of interests]{The authors report no conflict of interest.}

\backsection[Data availability statement]{The data that support the findings of this study are available from the corresponding author, O.T.S., upon reasonable request.}

\backsection[Author ORCIDs]{B. Yeung, https://orcid.org/0009-0004-8040-4853; O. Schmidt, https://orcid.org/0000-0002-7097-0235}


\appendix

\section{Computational grids}\label{sec:appMesh}
\begin{figure}
    \centering
    \includegraphics{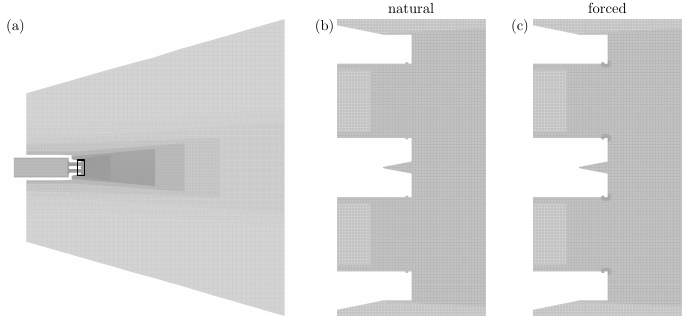}
    \caption{Computational grid along the major-axis plane, $y=0$. Panels (b) and (c), corresponding to the natural and forced cases, respectively, zoom in on the region in (a) marked by the black box.}
    \label{fig:mesh}
\end{figure}
Figure \ref{fig:mesh}(a) shows a cross section of the unstructured, Voronoi-based computational grid along the major-axis plane, $y=0$. The grid contains successive levels of refinement, from a coarse mesh in the far field to a fine mesh near the nozzle. The grids of the natural and forced cases are nearly identical, except near the nozzle lips. In and around the cavity that is immediately upstream of the nozzle lip, the forced case in figure \ref{fig:mesh}(c) halves the average cell width of the natural case in \ref{fig:mesh}(b). This additional refinement was implemented to improve the capture of the plasma actuation \citep{BresEtAl2021AIAA}, which is located inside the cavity.

\section{Effect of observable on recovery of SPOD from BMD}\label{sec:appBmd_recover_spod}
\begin{figure}
    \centering
    \includegraphics{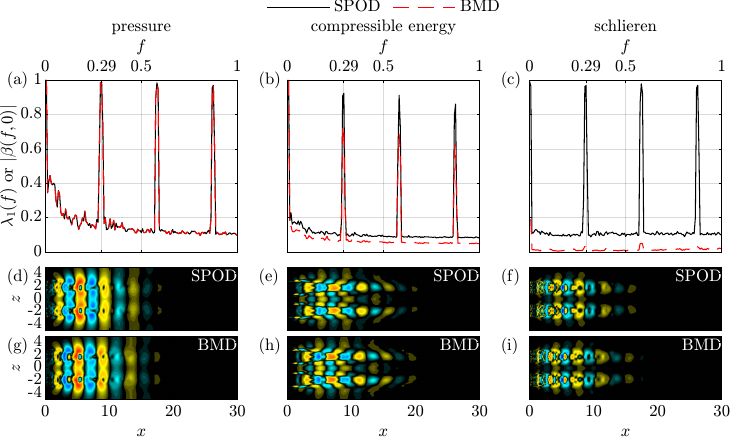}
    \caption{Comparison between the leading SPOD eigenvalue spectrum for SS symmetry, and BMD mode bispectrum along the $f_l=0$ axis for SS-SS interactions: (a) pressure 2-norm; (b) compressible energy norm; (c) numerical schlieren 2-norm. The data are normalised in accordance with \S \ref{sec:bicoherence}. For each norm, the leading SPOD mode at $f=f_0$ and bispectral mode at $(f_k,f_l)=(f_0,0)$ are shown in the second and third rows, respectively. In (e,h), the $u_x$ component of each mode is displayed.}
    \label{fig:bmd_recover_spod}
\end{figure}
In \S \ref{sec:bmd_recover_spod}, we stated that, if the mean flow is retained, the BMD performed on pressure data can recover the leading SPOD eigenvalues and modes along the abscissa, $f_l=0$, or ordinate, $f_k=0$. Using the SS symmetry as an example, figure \ref{fig:bmd_recover_spod}(a) demonstrates that the leading SPOD eigenvalue, $\lambda_1(f)$, is perfectly recovered by the magnitude of the BMD mode bispectrum along the abscissa, $|\beta(f,0)|$. While not shown here, we have also confirmed that the BMD of density and temperature similarly recovers the SPOD. In contrast, if the observable includes velocity, the non-uniformity of the mean velocity prevents the quantitative recovery of the SPOD. Figure~\ref{fig:bmd_recover_spod}(b) compares $\lambda_1(f)$ and $|\beta(f,0)|$ for the state vector $\vb*q=[\rho,u_x,u_y,u_z,T]^\mathrm{T}$ and the compressible energy norm \citep[][see also appendix \ref{sec:appNorm}]{Chu1965ActaMech}. Although the mode bispectrum deviates from the SPOD spectrum, the two remain qualitatively matched. For the state vector $\vb*q=[u_x,u_y,u_z]^\mathrm{T}$ (not shown), BMD performs similarly.

Experimental data collected from supersonic jets are often in the form of schlieren images. To investigate the applicability of mean-retained BMD to schlieren data, we compute the streamwise gradient of the LES density, $\pdv*{\rho}{x}$, which serves as a numerical schlieren. The comparison between SPOD and BMD of $\pdv*{\rho}{x}$ is reported in figure \ref{fig:bmd_recover_spod}(c). In this case, due to the highly non-uniform mean, $\pdv*{\bar\rho}{x}$, the bispectrum fails to capture the SPOD spectrum. Nevertheless, in all three cases---pressure 2-norm, compressible energy norm, and schlieren 2-norm---the SPOD and BMD modes appear identical, up to an arbitrary phase difference.

\begin{figure}
    \centering
    \includegraphics{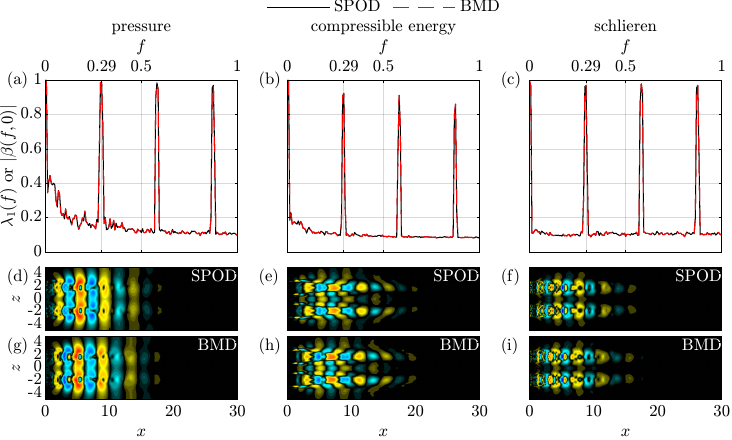}
    \caption{Same as figure \ref{fig:bmd_recover_spod}, but after setting the zero-frequency Fourier realisations to unity for the BMD.}
    \label{fig:bmd_unitmean_recover_spod}
\end{figure}
For arbitrary data, SPOD can always be perfectly recovered from BMD by replacing the zero-frequency Fourier realisations with unity. Figure~\ref{fig:bmd_unitmean_recover_spod} repeats the calculations in figure \ref{fig:bmd_recover_spod} but for this replacement. The BMD bispectrum now matches its corresponding SPOD spectrum for all three choices of norm. In the latest version of our \textsc{Matlab} implementation of BMD, available at \url{https://www.mathworks.com/matlabcentral/fileexchange/83408-bispectral-mode-decomposition}, we provide Fourier-mode replacement as an option for the user.

\section{Exploiting spatial symmetry in modal decompositions}\label{sec:appD2}
For axisymmetric jets, it is well-known that continuous rotational symmetry should be enforced on the modal decomposition via an azimuthal Fourier decomposition of the data \citep{Sirovich1987QAM2,BerkoozEtAl1993AnnuRev}. For discrete symmetries, some ambiguity exists in the literature. As described by \citet{Sirovich1987QAM2}, symmetry group operations can be used to inflate the ensemble size of the POD problem (and its variants, e.g. SPOD). The same technique may be applied to BMD. In the context of $D_2$ symmetry, for a given frequency $f_k$, we can define an inflated ensemble matrix,
\beq
\tilde{\hat{\mathsfbi{Q}}}_k(x,y,z) = \mqty[\hat{\mathsfbi{Q}}_k(x,y,z), \hat{\mathsfbi{Q}}_k(x,-y,z), \hat{\mathsfbi{Q}}_k(x,y,-z), \hat{\mathsfbi{Q}}_k(x,-y,-z)] \in\mathbb R^{n_\mathrm{dof}\times 4n_\mathrm{blk}},
\eeq
which is quadruple the size of $\hat{\mathsfbi{Q}}_k$. The four submatrices in $\tilde{\hat{\mathsfbi{Q}}}_k$ contain equivalent realisations of the Fourier transform. For the frequency pair $(f_k,f_l)$, the inflated bispectral density matrix is then given by
\beq
\tilde{\mathsfbi{B}}_{k,l} = \frac{1}{\nblk}{\tilde{\hat{\mathsfbi{Q}}}}_{k\circ l}^* \mathsfbi{W} {\tilde{\hat{\mathsfbi{Q}}}}_{k+l}.
\eeq
The numerical radius problem for $\tilde{\mathsfbi{B}}_{k,l}$ can be solved in the same manner as previously outlined in \S \ref{sec:method}.

Our approach, the $D_2$ decomposition, corresponds to a later insight by \citet{SirovichPark1990POF} that does not increase the ensemble size. Also developed originally for POD, the approach instead decomposes each realisation into components that possess perfect symmetry or antisymmetry. Just as the azimuthal Fourier decomposition is a weighted sum over the azimuth, the $D_2$ decomposition is a weighted sum over quadrants. The Fourier decomposition of round jets is thus closely related to the $D_2$ decomposition of twin-rectangular jets. An additional advantage of symmetry decompositions is that they enable symmetry-cross interactions to be examined using cross-BMD.

\section{Choice of norm}\label{sec:appNorm}
\begin{figure}
    \centering
    \includegraphics{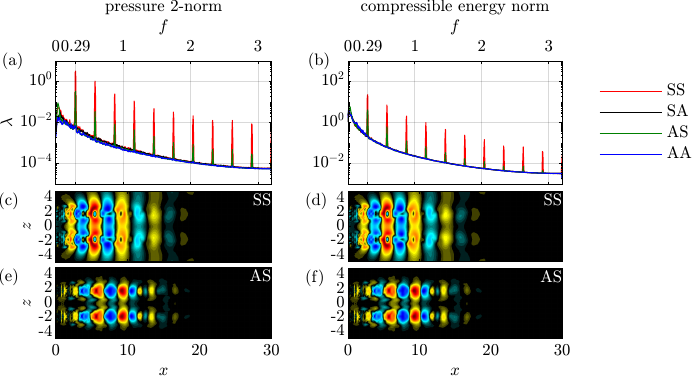}
    \caption{Leading SPOD eigenvalues (a,b) and modes at $f=0.29$ for the SS (c,d) and AS (e,f) symmetries, computed using the pressure 2-norm (left column) and compressible energy norm (right column). For the modes, the $y=0.25$ plane is shown.}
    \label{fig:norm}
\end{figure}
The SPOD and BMD problems in \S\S \ref{sec:energetics}--\ref{sec:nlin} are based on the pressure 2-norm. In this appendix, we justify this choice by comparing the SPOD eigenvalue spectra and modes of the twin jet based on the pressure 2-norm and compressible energy norm \citep{Chu1965ActaMech}. For the compressible energy norm, we define the state vector $\vb*q=\qty[\rho, u_x, u_y, u_z, T]$. The variables $\rho$, $u_x$, and $T$ are subjected to the same $D_2$ decomposition as $p$ in \S \ref{sec:d2}. The symmetry components of $u_y$ and $u_z$ are obtained as
\begin{IEEEeqnarray}{rCl}
u_{y,\mathrm{SS}} &=& (1/4)\qty[u_y(x,y,z,t) - u_y(x,-y,z,t) + u_y(x,y,-z,t) - u_y(x,-y,-z,t)] \IEEEeqnarraynumspace\IEEEyesnumber\IEEEyessubnumber\\
u_{y,\mathrm{SA}} &=& (1/4)\qty[u_y(x,y,z,t) - u_y(x,-y,z,t) - u_y(x,y,-z,t) + u_y(x,-y,-z,t)] \IEEEeqnarraynumspace\IEEEyessubnumber\\
u_{y,\mathrm{AS}} &=& (1/4)\qty[u_y(x,y,z,t) + u_y(x,-y,z,t) + u_y(x,y,-z,t) + u_y(x,-y,-z,t)] \IEEEeqnarraynumspace\IEEEyessubnumber\\
u_{y,\mathrm{AA}} &=& (1/4)\qty[u_y(x,y,z,t) + u_y(x,-y,z,t) - u_y(x,y,-z,t) - u_y(x,-y,-z,t)] \IEEEeqnarraynumspace\IEEEyessubnumber
\end{IEEEeqnarray}
and
\begin{IEEEeqnarray}{rCl}
u_{z,\mathrm{SS}} &=& (1/4)\qty[u_z(x,y,z,t) + u_z(x,-y,z,t) - u_z(x,y,-z,t) - u_z(x,-y,-z,t)] \IEEEeqnarraynumspace\IEEEyesnumber\IEEEyessubnumber\\
u_{z,\mathrm{SA}} &=& (1/4)\qty[u_z(x,y,z,t) + u_z(x,-y,z,t) + u_z(x,y,-z,t) + u_z(x,-y,-z,t)] \IEEEeqnarraynumspace\IEEEyessubnumber\\
u_{z,\mathrm{AS}} &=& (1/4)\qty[u_z(x,y,z,t) - u_z(x,-y,z,t) - u_z(x,y,-z,t) + u_z(x,-y,-z,t)] \IEEEeqnarraynumspace\IEEEyessubnumber\\
u_{z,\mathrm{AA}} &=& (1/4)\qty[u_z(x,y,z,t) - u_z(x,-y,z,t) + u_z(x,y,-z,t) - u_z(x,-y,-z,t)], \IEEEeqnarraynumspace\IEEEyessubnumber
\end{IEEEeqnarray}
respectively. We seek modes that are optimal under the inner product
\beq
\expval{\vb*q_1,\vb*q_2}_{\vb*x} = \int_\mathit\Omega \vb*q_2^*\operatorname{diag}\qty(\qty[\bar T/(\gamma\bar\rho),\bar\rho,\bar\rho,\bar\rho,\bar\rho/(\gamma(\gamma-1)\bar T)])\vb*q_1\dd{\vb*x}.
\eeq
Note that unlike in \citet{SchmidtEtAl2018JFM}, e.g., the Mach number does not appear in the weight tensor here. This is because the primitive variables are non-dimensionalised by the ambient conditions. Figure~\ref{fig:norm} reports the leading SPOD eigenvalues and modes based on each norm. The spectra in figure \ref{fig:norm}(a) based on the pressure 2-norm display tones at the fundamental forcing frequency, $f_0=0.29$, and its harmonics in the SS and AS symmetry components. In figure \ref{fig:norm}(b), the spectra based on the compressible energy norm present weaker tones in the same symmetry components. For these symmetries, the leading modes at $f_0$ for the pressure 2-norm, in figure \ref{fig:norm}(c,e), and the compressible energy norm, in figure \ref{fig:norm}(d,f), appear visually indistinguishable. Since no tones are observed in the SA and AA symmetries, their corresponding modes are omitted. In a systematic comparison of norms based on pressure or turbulent kinetic energy, for the space-only POD of a compressible turbulent jet, \citet{FreundColonius2009IJA} found that the pressure norm reconstructs pressure fluctuations more efficiently, and velocity fluctuations nearly as efficiently, as the latter. For the twin jet, it thus stands to reason that the pressure norm, while more restrictive than the compressible energy norm, nevertheless captures qualitatively the same dynamics. The nonlinear dynamics, in particular, are more active under the pressure norm, as figure \ref{fig:norm}(a,b) reveals. In our BMD analysis, we therefore follow the methodology of \citet{Schmidt2020NDyn}, and focus solely on pressure.




\bibliographystyle{jfm}
\bibliography{refs}


\end{document}